\pdfoutput=1
\documentclass[aps,pra,twocolumn,superscriptaddress,nofootinbib,floatfix,longbibliography]{revtex4-1}
\usepackage{amsmath}
\usepackage{latexsym}
\usepackage{amssymb}
\usepackage{bm}
\usepackage{color}
\usepackage[usenames,dvipsnames,svgnames]{pstricks} 
\usepackage{mathtools}
\usepackage{amsfonts}
\usepackage{amsthm}
\usepackage{graphicx}
\usepackage{hyperref}
\usepackage{tikz}
\newcommand{\rr}{\textcolor{black}}
\newcommand{\bb}{\textcolor{black}}

\begin{document}
\title{Landauer Principle and Thermodynamics of Computation}
\author{Pritam Chattopadhyay}
\thanks{Contributed equally}
\email{pritam.cphys@gmail.com}
\affiliation{Cryptology and Security Research Unit, Indian Statistical Institute, Kolkata 700108, India}
\affiliation{Department of Chemical and Biological Physics,
Weizmann Institute of Science, Rehovot 7610001, Israel}
\author{Avijit Misra}
\thanks{Contributed equally}
\email{avijitmisra@iitism.ac.in}
\affiliation{Department of Physics, Indian Institute of Technology (Indian School of Mines), Dhanbad, Jharkhand 826004, India}
\affiliation{Department of Chemical and Biological Physics,
Weizmann Institute of Science, Rehovot 7610001, Israel}

\author{Tanmoy Pandit}
\email{tanmoypandit163@gmail.com}
\affiliation{Hebrew University of Jerusalem, Jerusalem 9190401, Israel}
\affiliation{Department of Mathematics and Physics, Leibniz University of Hannover, Hanover 30167, Germany}
\affiliation{Institute for Theoretical Physics, Technical University of Berlin, Berlin 10623, Germany}

\author{Goutam Paul}
\email{goutam.paul@isical.ac.in}
\affiliation{Cryptology and Security Research Unit, Indian Statistical Institute,
Kolkata 700108, India}


\date{\today}

\begin{abstract}
According to the Landauer principle, any logically irreversible process accompanies entropy production, which results in heat dissipation in the environment. Erasing of information, one of the primary logically irreversible processes, has a lower bound on heat dissipated into the environment, called the Landauer bound (LB). However, the practical erasure processes dissipate much more heat than the LB. Recently, there have been a few experimental investigations to reach this bound both in the classical and quantum domains. There has also been a spate of activities to enquire about this LB in finite time, with finite-size heat baths, non-Markovian and nonequilibrium \rr{environments} in the quantum regime where the effects of fluctuations and correlation of the systems with the bath can no longer be ignored. This article provides a comprehensive review of the recent progress on the Landauer bound, which serves as a fundamental principle in the thermodynamics of computation.  We also provide a perspective for future endeavors in these directions.

Furthermore, we review the recent \rr{explorations} toward establishing energetic bounds of a computational process. We also \rr{discuss} the thermodynamic aspects of error correction, which is an indispensable part of information processing and computations. In doing so, we briefly discuss the basics of these fields to provide a complete picture.

\end{abstract}

\maketitle

\tableofcontents
\newpage
\noindent
\textbf{\Large Notations:}\\
\vspace{1ex}

\noindent
$\bullet$ $\mathbb{S}\,\,\, \epsilon \,\, \{0,1\}$\\
$\bullet$ $\mathbb{S}^\star \, \rightarrow$ the set of all binary strings.\\
$\bullet$
$\delta_{ij}=\left\{
\begin{array}{lc}
  1, &\  i = j,\\
  0, & \  \text{otherwise}.
\end{array}\right.$\\
$\bullet$ $\rho \rightarrow$ density matrix\\
$\bullet$ $H \rightarrow$ Hamiltonian\\
$\bullet$ $S \rightarrow$ Entropy\\
$\bullet$ $U \rightarrow$ Unitary\\
$\bullet$ $\dagger \rightarrow$ complex conjugate\\
$\bullet$ $\otimes \rightarrow$ tensor product\\
$\bullet$ $k_B \rightarrow$ Boltzmann constant\\
$\bullet$ $\beta \rightarrow \frac{1}{k_B T}$ \\
$\bullet$ $\text{Tr}$ (or\, $ \text{tr}$) $\rightarrow$ trace\\
$\bullet$ $\ln \rightarrow$ log$_e$\\
$\bullet$ $eV \rightarrow$ Electron volt\\
$\bullet$ J $\rightarrow$ Joule\\

\noindent
\textbf{\Large Acronym:}\\

\noindent
$\bullet$ RNA $\rightarrow$ Ribonucleic Acid\\
$\bullet$ TTL $\rightarrow$ Transistor Transistor Logic\\
$\bullet$ MD $\rightarrow$ Maxwell\rr{'s} Demon\\
$\bullet$ LB $\rightarrow$ Landauer Bound\\
$\bullet$ LL $\rightarrow$ Landauer Limit\\
$\bullet$ LP $\rightarrow$ Landauer\rr{'s} Principle\\
$\bullet$ LE $\rightarrow$ Landauer Erasure\\
$\bullet$ OQS $\rightarrow$ Open Quantum System\\
$\bullet$ CGF $\rightarrow$ Cumulant Generating Functions\\
$\bullet$ TLS $\rightarrow$ Two-level System\\
$\bullet$ NM $\rightarrow$ Non Markovian \\
$\bullet$ CPTP $\rightarrow$ Completely Positive Trace Preserving\\
$\bullet$ ABEL $\rightarrow$ Anti-Brownian Electrokinetic\\
$\bullet$ MOKE $\rightarrow$ \rr{Magneto}-Optical Kerr Effect\\
$\bullet$ NMR $\rightarrow$ Nuclear Magnetic Resonance\\
$\bullet$ BLC $\rightarrow$ Ballistic Computer\\
$\bullet$ BWC $\rightarrow$ Brownian Computer\\
$\bullet$ TM $\rightarrow$ Turing Machine\\
$\bullet$ FA $\rightarrow$ Finite Automata\\
$\bullet$ FSM $\rightarrow$ Finite State Machine\\
$\bullet$ KC $\rightarrow$ Kolmogorov complexity\\
$\bullet$ DFA $\rightarrow$ Deterministic Finite Automata\\
$\bullet$ NFA $\rightarrow$ Non-deterministic Finite Automata\\
$\bullet$ PFA $\rightarrow$ Probabilistic Finite Automata\\
$\bullet$ ID $\rightarrow$ Instantaneous Description\\
$\bullet$ UTM $\rightarrow$ Universal Turing Machine\\
$\bullet$ EC $\rightarrow$ Error Correction\\
$\bullet$ ECC $\rightarrow$ Error Correcting code\\
$\bullet$ QEC $\rightarrow$ Quantum Error Correction\\
$\bullet$ LHS $\rightarrow$ Left Hand Side\\
$\bullet$ RHS $\rightarrow$ Right Hand Side\\
$\bullet$ MOS $\rightarrow$ Metal Oxide Semiconductor\\ 
$\bullet$ CMOS $\rightarrow$ Complementary Metal Oxide Semiconductor\\
$\bullet$ RTO $\rightarrow$ Restore To One\\
$\bullet$ BHT $\rightarrow$ Brassard Hoyer Tapp\\
$\bullet$ RAM $\rightarrow$ Random Access Memory\\
$\bullet$ QPT $\rightarrow$ Quantum Phase Transition\\

\section{Introduction}
\bb{John von Neumann, in his 1948 lectures, posed a fundamental question \rr{on} whether a computer operating at temperature $T$ must necessarily dissipate heat. The details of this lectures were later formalized in his book \textit{Theory of Self-Reproducing Automata}, completed in 1966 by Arthur W. Burks~\cite{von1966theory}. \rr{In 1961,} Landauer asserted that “\textit{real-world computation involves thermodynamic costs}” and emphasized the implications of \rr{the same}~\cite{landauer1961irreversibility,landauer1991information}.}
\bb{ The \textit{man-made computers} and even all naturally occurring processes like \textit{biological computers} have thermodynamic costs.} It is quite fascinating to analyze the difference in the thermodynamic cost of the naturally occurring process and the artificial ones that are created. Translation of ribonucleic acid (RNA) into amino acids is one such natural biological process where one encounters energy costs for the execution of the process. \rr{These biological processes are thermodynamically more efficient than the artificial ones~\cite{kempes2017thermodynamic}.} 

Among the myriad artificial processes engineered by humans, digital computation stands out as one of the most significant. Modern digital computers can be viewed as engines that irreversibly dissipate energy to execute mathematical and logical operations. Early scientific thought postulated that there must exist a fundamental thermodynamic bound on the efficiency of such computational engines, independent of the hardware architecture employed. However, contemporary understanding has revealed a more refined reality: while the fundamental thermodynamic limit for the energy cost of erasing a single bit is set by $k_B T \ln 2$ (where $k_B$ is the Boltzmann constant and $T$ is the operational temperature), modern computers dissipate energy per logic operation that exceeds this bound by many orders of magnitude. Consequently, despite being capable of executing a vast number of reliable computations, practical devices remain highly inefficient relative to the theoretical minimum. A principal cause of this inefficiency lies in the reliance on volatile memory elements—such as Transistor-Transistor Logic (TTL) flip-flops~\cite{chandrakasan1992low,horowitz20141}—which inherently waste energy.
The macroscopic nature of the existing computers is one of the basic reasons for the inefficiency in the context of energy. One of the spectacular thermodynamically reversible computation models is the ballistic \rr{computer}, proposed by Fredkin and Toffoli~\cite{fredkin1982conservative}. \rr{However, other models~\cite{bennett1973logical,keyes1970minimal,likharev1982classical} have been developed, which are} more physically realistic than that of Fredkin's version.

These examples of artificial and natural systems hinge on the deep connection between computation and thermodynamics. \rr{The connection between thermodynamics and a logically irreversible process gets most prominent in the context of  \textit{Maxwell's demon} (MD)~\cite{maxwell1871theory,maruyama2009colloquium,junior2025friendly}.  Landauer (later Bennett) argued that one must pay an entropic cost that gets dumped into the environment as heat while performing a logically irreversible process that erases or throws away information. This argument played the pivotal role in exorcising the MD and thereby saving the second law of thermodynamics. It also established that \textit{information is physical.} There have been several explorations to model MD in physical systems and study Landauer's principle (LP). However, they mostly belong to \textit{equilibrium statistical physics}. But when we approach to the miniaturized domain,  the systems are, in general, out of equilibrium.} The major breakthroughs in \textit{non-equilibrium statistical physics}~\cite{zwanzig2001nonequilibrium,kubo2012statistical,PhysRevLett.78.2690,PhysRevE.56.5018} allow us to analyze the thermodynamic behavior of systems that are arbitrarily far from their equilibrium, and even for systems that undergo arbitrary external driving.  With the advent of non-equilibrium statistical physics, the researchers have shown a keen interest to analyze the thermodynamic cost for the erasure process in this domain which includes finite time, finite size, \rr{non-Markovian (NM) open quantum dynamical effects}, and so on~\cite{goold2015nonequilibrium, khoudiri2025non,lorenzo2015landauer,proesmans2020finite,miller2020quantum,esposito2011second,mandal2012work,helms2022stochastic,ray2023gigahertz,kuang2022modelling,proesmans2021erasing,barnett2013beyond,berta2018deconstruction,henao2023catalytic,barato2013autonomous,mandal2013maxwell,deffner2013information,barato2014stochastic,strasberg2014second,PhysRevE.106.034112,mondal2023modified}. 

\bb{In nature, copying information is a fundamental mechanism within natural systems~\cite{watson1953molecular,hopfield1974kinetic,ouldridge2017thermodynamics,horowitz2017spontaneous}. Yet, replication is inherently prone to error.} The accuracy of a copy thus relies on its accurate reproduction. It can be quantified by counting the wrongly copied bits while execution. Though the error can be reduced at the macro level, but at the molecular level, perfect copying is not achievable due to thermal fluctuations. It is the primary source of error at the molecular level. The replication process is limited by the thermal noise, so it must be interpreted in terms of thermodynamics as proposed by von Neumann~\cite{von1956probabilistic}. This raises an important question: \textit{Can one establish a fundamental connection between the thermodynamic cost and the errors incurred during the information copying process}?  

Generally, while executing a copying process, it has to undergo various intermediate steps to control the accuracy and speed of the process. This is true for artificial as well as natural scenarios~\cite {johnson1993conformational}.  \rr{If we try} to explain the errors from thermodynamic laws (the second law), \rr{one must account for the fact that the copying process is repeated cyclically rather than occurring as a \textit{single-shot} operation}~\cite{bennett1982thermodynamics}.  
\rr{The works~\cite{sartori2015thermodynamics, korepin2002thermodynamic} forge a profound link between thermodynamics and informational errors, demonstrating that errors arising in copying protocols are intrinsically connected to thermodynamic observables, which characterize the errors.} 

\rr{Understanding the thermodynamic cost of computation is essential in both classical and quantum computation, as it reveals how the physical laws fundamentally constrain computational efficiency. Core models such as \textit{reversible computation}, \textit{Turing machines}, and \textit{finite automata} offer deep theoretical insights into the thermodynamic cost of computational models. Reversible computing frameworks, in principle, allow for vanishingly low energy dissipation by preserving information throughout the process, yet practical limitations inevitably lead to non-negligible thermodynamic costs. Similarly, reversible Turing machines can asymptotically approach this ideal limit, though operations such as memory resetting and tape erasure introduce unavoidable energy losses. Finite automata, despite their relative simplicity, capture the essence of how logical state transitions and memory utilization contribute to entropy production. Collectively, these models provide a rigorous foundation for exploring the intrinsic trade-offs among energy consumption, computational speed, and information preservation in physically realizable computing systems.}

 \rr{In this article, we review the recent developments on LP in various contexts, including the most recent experimental demonstrations, thermodynamic aspects, and cost analysis of computational models and error correction.}

\vspace{2ex}
\begin{center}
\textbf{Scope of the review}
\end{center}
\vspace{2ex}
\rr{Aiming to highlight recent advancements and fundamental insights into LP and the thermodynamics of computation, this review addresses several key aspects spanning both theoretical and experimental fronts. We begin in Sec.~\ref{sec2} with a discussion of the generalized LP. Sec.~\ref{sec.3} reviews significant recent developments on LP, including scenarios involving finite time dynamics, finite-size and non-Markovian reservoirs. In Sec.~\ref{sec4}, we explore the role of Landauer limit (LL) in computational processes, followed by Sec.~\ref{sec.5}, which presents an overview of recent experimental efforts aimed at realizing and validating LL in practical systems.}

\rr{Consequently, we move towards exploring the thermodynamic ramifications and implications for computation and error correction. Sec.~\ref{sec.6} delves into reversible computational models, laying the groundwork for a broader discussion in Sec.~\ref{sec7} on the thermodynamics of computational paradigms, including finite state automata and Turing machines. In Sec.~\ref{sec8}, we discuss the thermodynamic consistency of error-correcting codes. Prior to concluding and outlining potential future research directions in Sec.~\ref{sec10}, we briefly address a range of complementary topics such as viewing the computer as a heat engine, the thermodynamics of algorithms, and the Landauer bound (LB) in switching protocols, under the umbrella of miscellaneous discussions in Sec.~\ref{sec9}.}

\vspace{2ex}
\begin{center}
\textbf{Areas not covered in this review}
\end{center}
\vspace{2ex}
\rr{Though the review covers an extensive amount of works in the context of LP and thermodynamics of computation, there are some related topics and work, \bb{that are not discussed here, but are worth mentioning}. For example, while most of the aspects of the LP have been reviewed here,  some of the left-out aspects in this domain are like the role of LP in gravity~\cite{bormashenko2019landauer,haranas2021landauer,daffertshofer2007forgetting},  relativity~\cite{herrera2020landauer}, quantum field theory~\cite{xu2022landauer}, many-body phenomena~\cite{bonancca2023information,parrondo2001szilard}, material~\cite{zivieri2022thermodynamics}.}

\rr{In this review, we have considered mainly two fundamental computational models: the Turing machine and the Finite state machine, namely, for detailed thermodynamic analysis. Some aspects, like LB in analog computers and algorithmic thermodynamics, have not been discussed here. Interested readers can go through~\cite{diamantini2016landauer} for the LB analysis in the analog system and~\cite{baez2012algorithmic} for algorithmic thermodynamics, which in turn allows one to apply the laws and techniques of thermodynamics for studying algorithmic information theory; however, we have touched upon the thermodynamic cost of algorithms briefly in Sec.~\ref{sec9}.}

\rr{In what follows, we briefly deal with computational complexity when we estimate the energetic cost of computation. However, a somewhat related direction, the fundamental limitation on the computability of the physical process, is not discussed in this review. Interested readers can have a look at the following seminal articles~\cite{pour1982noncomputability,moore1990unpredictability,lloyd2000ultimate,lloyd2017uncomputability}. Similarly, the thermodynamics of controlled systems is not covered. Interested readers can go through~\cite{touchette2004information,touchette2000information,barato2017thermodynamic,sagawa2008second,sagawa2012nonequilibrium,wilming2016second,large2021stochastic,gingrich2016near,horowitz2017information} for details.}

 \rr{ Another important aspect that has not been covered here is the thermodynamic analysis of biological and biochemical processes. For the same, one can go through~\cite{ouldridge2017fundamental,ouldridge2018importance,brittain2019biochemical,sartori2014thermodynamic,hasegawa2018multidimensional,mehta2012energetic,mehta2016landauer,lan2012energy,ouldridge2017thermodynamics,govern2014optimal,barato2015thermodynamic}. Similarly, we do not cover the modeling of computational machines based on biochemical and biological systems. Interested readers can go through ~\cite{prohaska2010innovation,bryant2012chromatin,benenson2012biomolecular,chen2014deterministic,dong2012bisimulation,soloveichik2008computation,mougkogiannis2025response} to have a clear idea in this direction.}

\section{Landauer's Principle}\label{sec2}
The search for the development of more impactful computing circuits leads to the question: \textit{What would be the physical limitation of this process?}  Rolf Landauer, in the year 1961, in his seminal work~\cite{landauer1961irreversibility} proposed an important limit to the \rr{posed question} by von Neumann~\cite{von1966theory}. Landauer stated that an irreversible process, \rr{where information is either deleted or thrown away,}   dissipates a minimum amount of  \rr{energy per bit}~\cite{bennett2003notes}, \rr{which is dumped as heat into the environment, increasing its entropy.} \rr{This minimum amount of heat dissipated to erase one bit of information, known as ``\textit{Landauer bound}" (LB), is given by}
\begin{equation}\label{LPeq1}
   Q \geq k_B T \ln 2,
\end{equation}
\bb{where $k_B$ is the Boltzman constant} and $Q$ is \rr{the heat dumped into the environment at temperature $T$.} The bound at room temperature is approximately 0.018 $eV$ (2.9 x $10^{-21}$ J), whereas the modern world computers use a \textit{billion times} more energy~\cite{PhysRevLett.107.010604,moore2012landauer}.

\rr{In 1867, Maxwell proposed a thought experiment, which he later elaborated on in his book {\it ``Theory of heat''}~\cite{maxwell1871theory}, where a demon has a chamber of ideal gas in equilibrium at some fixed temperature. The demon creates a partition between the chamber and allows the molecules to cross the partition from, say, left to right if their velocity is greater than some threshold value. The molecules in the right can only go to the left if their velocities are less than the threshold value. Thus, the demon can create a temperature difference, which can further be exploited to extract work in principle, but without performing any work. This contradicts the second law of thermodynamics, as one can extract work from a single temperature source.  Szilard proposed a simplified version of this thought experiment with a single gas molecule known as  Szilard's engine in 1929. This thought experiment puzzled the scientific communities for almost a century till the discovery of LP in 1961~\cite{landauer1961irreversibility}. The exorcism of MD can be done by virtue of the fact that the demon needs at least the amount of energy it has extracted earlier, on average, to erase their memory, before it can work in the next cycle. Brillouin's argument to reconcile MD with the second law was that the demon must pay an entropic cost to acquire information, which he dubbed as negentropy~\cite{brillouin1962science}. It was Bennet who showed in 1982~\cite{bennett1982thermodynamics} that the information gain and storage can be done reversibly in principle, and confirmed that only the erasure step must bear the entropic cost. Thus, the LP successfully exorcises the MD.}

 Though widely accepted for reconciling MD with the second law~\cite{ladyman2007connection,bennett2003notes}, Landauer's Principle (LP) has faced various criticisms and objections~\cite{earman1998exorcist,earman1999exorcist,shenker1998maxwell,maroney2005absence,norton2005eaters,norton2011waiting,bennett2003notes}. The prime objections to LP were put forth by Earman and Norton~\cite{earman1999exorcist}. They argued that LP is dependent on the second law of thermodynamics; it can be considered to be either unnecessary or insufficient for ``\textit{exorcism of Maxwell’s Demon}". \rr{In~\cite{sagawa2008second,cao2009thermodynamics} it has also been argued recently that LP is a consequence of the second law.} The other objections over LP are generally of three kinds~\cite{bennett2003notes}:
(a) \rr{According to many, at that time at least,} thermodynamic quantities such as heat and work are fundamentally unrelated to mathematical concepts like logical irreversibility. \rr{Therefore}, drawing direct parallels between them \rr{seems unfounded}.  
(b) In all cases of data-processing operations, there is a dissipation of at least $k_B T \ln 2$ amount of energy irrespective of the condition whether it is logically reversible or not. \rr{In a realistic physical scenario, it is even much more.}
(c) Another reason for objection is that, in principle, \rr{even} logically irreversible operations can be engineered in a thermodynamically reversible way. \rr{In~\cite{bennett2003notes}, the author discussed in detail and refuted all these criticisms.}

Attempts have been made to \rr{establish} LP \rr{rigorously}. Piechocinska~\cite{piechocinska2000information} offered \rr{a} proof of LP with the help of statistical mechanics. The proof \rr{assumes} a particular physical model. The notable advancements \rr{have been done} in~\cite{turgut2009relations,ladyman2007connection,ladyman2008use,cao2009thermodynamics,leff2002maxwell}, where \rr{LP has been generalized without assuming a particular physical model.}
\rr{ A notable} generalization of the LP surfaced in~\cite{vaccaro2011information}, where it has been shown that the information erasure will cause an increase in the entropy of the \rr{environment} with no energy cost, but the cost can be \rr{attributed to} the angular momentum of \rr{a spin-reservoir. In fact, angular momentum is exploited rather than energy from a spin-reservoir to erase the memory, and it is consistent with the second law of thermodynamics}.  

\rr{With rapid technological advancements and our ever increasing grasp on the miniaturized scale, probing and controlling the small scale systems that exhibits quantum traits are becoming more and more possible.}
\rr{In this regime, where we can no longer ignore the fluctuations, quantum correlations between the system and the environment}, the quantum effects \rr{bring fundamental challenges} in the analysis of memory erasure.  The validity of the erasure principle in this limit has been questioned~\cite{allahverdyan2000extraction,nieuwenhuizen2002statistical,horhammer2005thermodynamics,hilt2009system}. It has been claimed that due to the presence of entanglement between the system and the environment LP is \rr{violated}, which implies information is erased while heat is absorbed~\cite{allahverdyan2001breakdown,horhammer2008information,capek2005challenges,maruyama2009colloquium}. 
\bb{Hilt et al.~\cite{hilt2011landauer}, \rr{have} tackled this quantum conundrum and demonstrated that LP remains valid regardless of the specific nature of the interaction between the system and its environment. Their findings \rr{have} provided significant evidence supporting the applicability of LP in the quantum settings.}


To address the need for a theory with minimal assumptions, approaches \rr{have been} considered that derive the LP without relying explicitly on the second law of thermodynamics \rr{in both classical and quantum regimes}~\cite{shizume1995heat,piechocinska2000information,sagawa2009minimal,sagawa2011erratum}. Reeb \rr{and Wolf's} work~\cite{reeb2014improved} is \rr{the most significant one along this direction in quantum setups}.  They \rr{have} considered \rr{the following minimal set of assumptions}:
(a) In the process, the system and the reservoir belong to \rr{corresponding} Hilbert spaces.
(b) The reservoir is initially in a thermal state.
(c) The system and the reservoir \rr{are} initially uncorrelated. 
(d) The evolution process is unitary \rr{on the joint system-resrvoir state}.

With this minimal setup, a sharpened equality version of LP has been derived for a system $\mathcal{S}$ and a reservoir $R$, \rr{for the dissipated heat $Q$ to the environment} as 
\begin{equation}\label{LPeq2}
    \beta Q = \Delta S + I (\mathcal{S}^\prime : R^\prime) + S (\rho_R^\prime|| \rho_R)\geq \Delta S.
\end{equation}
Here $\prime$ denotes the final states, \rr{$\Delta S$ is the decrease in von Neumann entropy of the system}, $I (\mathcal{S}^\prime: R^\prime)$ describes the mutual information \rr{which} quantifies the correlation of the system with the bath~\cite{nielsen2002quantum}, and $\rho_R$ denotes the reservoir state. \rr{$\beta= 1/k_BT$ denotes the inverse temperature of the bath}. $ S (\rho_R^\prime|| \rho_R)$ quantifies the free energy increase in the bath, where \rr{$S (\rho || \rho^\prime) = \text{Tr}\, [\rho \,\ln \, \rho] - \text{Tr}\, [\rho \,\ln \, \rho^\prime]$ is the quantum relative entropy between  $\rho, \,\rho^\prime$~\cite{nielsen2002quantum}.} 

They \rr{further} generalized \rr{the setup}~\cite{reeb2014improved} by considering an initial correlation in the process, \rr{which modifies} the assumption as \rr{follows}: 
(a) In the process, the system, reservoir, and memory are in a joint quantum state initially,
(b) the \rr{marginal state of the} reservoir is initially thermal,
(c) the heat \rr{and entropy} exchange are expounded on the marginal states, and
(d) the evolution is processed by an unital positive trace-preserving map. 
\bb{In this generalized setup, where the system, reservoir, and memory are initially correlated, the standard Landauer bound can be modified significantly. Initial quantum or classical correlations can effectively reduce the minimal heat dissipation required for information erasure. Since entropy and heat are evaluated from marginal states, part of the entropy change may be absorbed by shared correlations rather than dissipated as heat. When the evolution is governed by a unital, trace-preserving map, which does not reduce entropy on its own, the role of initial correlations becomes even more prominent.}

Even with the minimal assumption of Reeb \rr{and Wolf}, it is \rr{considered that the system is governed by non-unitary} dynamics. \bb{However, it is a difficult task to explore erasing in pure Hamiltonian dynamics for the system.} It has been conveyed in~\cite{holtzman2021hamiltonian} that it is possible to implement a \rr{bit-erasure} with no thermodynamic cost using Hamiltonian dynamics if one has the information of the system with infinite accuracy. In this case, it corresponds to the energy of the system being known with infinite accuracy. \rr{This requires diverging computational resources as we will discuss later in Sec.~\ref{sec.3} B.}


So far, we have discussed LB for a non-zero finite temperature $T$ of the environment. \textit{What if the temperature of the environment tends to zero?} In this limit (i.e., $T\rightarrow 0$) the bound in Eq.~\eqref{LPeq1} is trivial. It states that $Q \geq 0$ when $T=0$ as the bath is in the ground state. A non-trivial bound for this condition has been proposed in~\cite{timpanaro2020landauer}. There, the authors have assumed that initially the environment is in a thermal state, as in the original case of LP. The state of the system or the type of system-environment interaction is considered to be general. The improved bound is always greater than Eq.~\eqref{LPeq1} and coincides with it when $T$ is high. It is derived based on two principles, the positivity of mutual information and the maximum entropy principle~\cite{guiasu1985principle,wu2012maximum,presse2013principles}, namely, and can be expressed in terms of the equilibrium heat capacity of the bath. It is interesting to explore what the modifications will be on the bound when the system and environment are initially correlated for the limit $T\rightarrow 0$.  

\section{Landauer Principle considering the open-system dynamics}\label{sec.3}

\rr{The study of the dynamics of a quantum system that interacts with other system(s) (environment) is described as an open quantum dynamics~\cite{breuer2002theory,rivas2012open,rotter2015review,viola1999dynamical,breuer2016colloquium,de2017dynamics,mukhopadhyay2017dynamics,bhattacharya2017exact}. It plays a crucial role in almost every aspects of quantum technology as the system which we are interested in is affected by the surrounding noise~\cite{breuer2002theory}.}
 
 \rr{ Till now, we have discussed the LB considering the entropic bounds sans the dynamic detail or fluctuations during the erasing processes. The dynamics of erasing bear significant importance in practical scenarios. For example, LB is reached in infinite time. But the practical uses demand finite-time erasure. Also, during the erasing procedure, the environment can get far from equilibrium (due to finite size), and the fluctuations due to it are crucial for LB. \textit{What do the NM features add to the LP?} These questions are not only practically important, but can also bring forth fundamental issues. In the following, we review the progress on the finite-time, non-equilibrium, and non-Markovian erasing processes, respectively, mainly in the quantum domain.}

\subsection{Finite Time Landauer's Principle}

The heat dissipated in the environment to erase one bit of information is achieved \bb{under the assumption of a quasistatic transformation}. In practical scenarios, the information erasure occurs in a finite time. \rr{Therefore, the finite-time analysis~\cite{andresen2011current} bears much importance both in the classical and quantum regimes. There has been an upsurge of research on finite-time erasure of bits both in the quantum and \textit{stochastic thermodynamics}~\cite{seifert2012stochastic,van2015ensemble}. In this finite-time regime}, the erasing process takes place under non-equilibrium conditions, and the fluctuations in the dissipated heat become significant. This bears important consequences when \textit{designing minuscule logical devices} that must be able to combat destructive fluctuations that lie well above the LB.

Research on minimizing the average dissipation of a mesoscopic thermodynamic system during finite-time transformations has primarily focused on optimizing a limited (and often small) number of control parameters that influence the system's potential landscape~\cite{schmiedl2007optimal,bonancca2014optimal,sivak2012thermodynamic,tafoya2019using,plata2020finite,bryant2020energy,boyd2018thermodynamics,riechers2020balancing,rolandi2023finite}. A significant advancement in this domain was made by Aurell et al.~\cite{aurell2011optimal,aurell2012refined}, who, using stochastic thermodynamics, developed protocols with full control over the potential landscape to minimize entropy production in both slow and fast limits, while constraining the final state to a fixed microscopic probability distribution. Building on this approach, the authors in their work~\cite{proesmans2020finite} proposed a framework \rr{that enables full control over the system's potential without requiring a pre-specified final state, allowing more flexible and efficient erasure protocols. Using this, they derive tight lower and upper bounds on the dissipated work during optimal finite-time bit erasure.} Extensions~\cite{proesmans2020optimal} \rr{in the direction of~\cite{proesmans2020finite}} and alternative approaches~\cite{zhen2021universal,dago2021information,dago2022dynamics} in this direction, without accounting for quantum effects, have been explored to establish an optimal bound on the cost of erasure.

\bb{Reeb and Wolf’s seminal work~\cite{reeb2014improved} provided a rigorous generalization of LP, demonstrating that quantum coherence fundamentally alters the thermodynamic cost of erasure. Their analysis showed that when the erased state exhibits coherence in the energy eigenbasis, the dissipation cost necessarily exceeds the classical LB. This correction arises because coherence prevents full thermalization through classical energy exchange alone, requiring additional dissipation mechanisms.} 

\bb{Expanding on this, Miller et al.~\cite{miller2020quantum} investigated finite-time erasure under Markovian dynamics generated by the adiabatic Lindblad equation: $\dot{\rho}_t=\mathcal{L}\rho_t$.} The generator $\mathcal{L}$ satisfies quantum details balance condition with respect to the instantaneous control Hamiltonian $H_t$ that guarantees an instantaneous fixed point of the dynamics, $\pi_t=e^{-\beta H_t}/\mbox{Tr}(e^{-\beta H_t})$ such that $\mathcal{L}(\pi_t)=0$. Additionally, the driving of the control Hamiltonian is performed slowly in the time interval $t\in [0,\tau]$, relative to the relaxation timescale of the dynamics which implies that the system remains close to the equilibrium state at all times, i.e., $\rho_t= \pi_t+\tau^{-1} \delta \rho_t$ where corrections of higher orders $O(\tau^{-n})$ can be neglected. In this quasistatic limit, the dissipation approaches the LB. 


The LP is performed by taking the initial Hamiltonian $H_0 \simeq 0$  and then slowly increasing the energy gap of the Hamiltonian until it reaches far beyond $k_BT$. This is equivalent to ensuring the boundary conditions on the system's state: $\rho_0=\pi_0 \simeq I/d$ \bb{where $d$ denotes the dimensionality of the Hilbert space} and $\rho_\tau=\pi_\tau \simeq |0\rangle\langle 0|$. To obtain the full statistics of the dissipated heat, the authors have used the cumulant generating functions (CGF) and quantified the excess stochastic heat in addition to the LB as follows
\begin{equation}\label{Finitetime1}
    Q= k_BT \ln d+ Q_d+ Q_c.
\end{equation}
In \eqref{Finitetime1}, the quantities are averaged over many trajectories, which represent each run of the experiment. Here, $Q_d$ is the classical (diagonal) contribution and $Q_c$ is the coherent contribution in the dissipated heat excess to LB, both being non-negative. The classical heat distribution   $Q_d$ follows a Gaussian distribution like a classical process in the slow-driving limit~\cite{PhysRevResearch.2.023377}, whereas for the coherent case, due to the presence of non-negative higher order cumulants, $Q_c$ can be non-Gaussian.

Further, the authors of~\cite{miller2020quantum} demonstrate the fluctuations in dissipated heat with a two-level system (TLS) described by the following Hamiltonian 
\begin{equation}
H_t= \frac{\epsilon_t}{2}(\cos \theta_t \sigma_z+\sin \theta_t \sigma_x),
\end{equation}
which can well approximate the low-energy dynamics of a system in a double-well potential~\cite{RevModPhys.59.1}. \bb{Here $\epsilon_t$ is energy splitting, $\theta_t$ denotes the mixing angle, and $\sigma_i$ ($i=x,y,z$) represents the Pauli spin matrices.}
The thermal dissipation is realized by an adiabatic Lindblad master equation~\cite{albash2012quantum} in the slow driving and weak coupling (to a bosonic heat bath)  limit. The competition between the energetic ($\sigma_z$) and coherent tunneling is depicted by the mixing angle $\theta_t$. $\dot{\theta}_t=0$ describes the classical bit whereas $\dot{\theta}_t\neq0$ describes the non-commuting quantum double-well case. 

\rr{It is important to mention that perfectly cooling a system to its desired final state and perfect erasing of information are fundamentally equivalent~\cite{PRXQuantum.4.010332}, as both correspond to reducing the rank of the system to $1$. However, \textit{Nearst's unattainability principle}, often referred to as the \textit{third law of thermodynamics}, states that to perfectly cool a physical system, one needs \textit{diverging resources}, either in terms of energy or time. The authors of~\cite{PRXQuantum.4.010332} showed that it is possible to cool perfectly with finite energy and time provided one is equipped with coherent control whose complexity diverges. In essence, under coherent control, perfect cooling is possible if either time, energy, or control complexity diverges. They further showed that with incoherent control (or cooling aided by a heat engine), perfect cooling remains fundamentally unattainable within finite time and finite control complexity, regardless of the amount of energy extracted from the auxiliary engine.}

\rr{Consequently, they established that for cooling under incoherent control with finite control complexity, it is possible to cool a system arbitrarily close to a desired temperature within accuracy $\epsilon_\beta >0$, with infinite number of operations and thereby in infinite time, with a energy, as measured by the amount of heat drawn from a hot bath at inverse temperature $\beta_h$, which is arbitrarily close to the bound called    \textit{Carnot-Landauer limit}, that is} 
\rr{\begin{equation} 
    \left| \Delta E_H - \frac{1}{\eta_c} \Delta F_S^{(\beta)} \right| < \epsilon_E,
\end{equation}}
\rr{where $\eta_c = 1 - \frac{\beta_H}{\beta}$ is the Carnot efficiency, $\Delta E_H $ quantifies the energy drawn from the hot bath, and $\Delta F_S^{(\beta)}$ represents the free energy difference between the initial and final states of the system at inverse temperature $\beta$, and $\epsilon_E> 0$. The \textit{Carnot-Landauer bound} $\frac{1}{\eta_c} \Delta F_S^{(\beta)}$ reduces to the LB when $\beta_H=0$. Both these bounds can be reached asymptotically with diverging control complexity or time.}

\bb{The role of \rr{quantum features like} coherence in finite-time processes becomes even more pronounced when considering \textit{many-body effects} and \textit{collective erasure}. The collective effects of many-body systems in finite-time erasure~\cite{PhysRevLett.131.210401} demonstrated that many-body interactions can significantly reduce dissipated work in finite-time thermodynamic processes. Unlike independent qubit erasure, where excess work scales linearly with system size, many-body protocols exhibit sublinear scaling: $  W_{\text{diss}} \propto N^x, \quad x < 1$,  indicating an accelerated convergence to LB and an enhancement in the efficiency of information erasure.}



\bb{\rr{In this direction}, Van Vu and Saito~\cite{van2022finite} further examined the interplay between coherence and erasure speed, establishing that finite-time erasure introduces additional cost \rr{dependent on the distance error to the desired final state} \rr{and inversely proportional to the total time taken by the process}, beyond the standard LB. Their findings reinforced the idea that quantum coherence consistently amplifies dissipation, regardless of the control protocol or driving speed. These insights collectively deepen our understanding of the thermodynamic constraints on quantum information processing, emphasizing the \textit{inescapable energy costs} imposed by coherence, speed, and many-body effects.}

\rr{Importantly, one can implement an arbitrarily fast erasure process through a highly optimized algorithm that prescribes the exact microstate manipulations needed to transform any initial state into a desired pure final state. If algorithmic complexity is unconstrained, meaning we are allowed arbitrarily complex and precise instructions (programs), then it is in principle possible to design a perfect erasure 
protocol that acts optimally on any input, driving the system to a unique pure state in finite time. These algorithms would encode a complete understanding of the initial conditions and the exact 
transformations required, bypassing the usual thermodynamic cost that arises from ignorance or randomness in the input.  However, this does not violate LP; rather, it circumvents the physical 
cost by assuming unbounded computational resources, shifting the ``cost" from thermodynamic to computational complexity ~\cite{PhysRevA.40.4731}.}

\rr{In practice, limitations arise due to thermodynamic speed limits~\cite{deffner2017quantum} and coherence-induced excess dissipation~\cite{faist2015minimal}, finite computational resources.} \rr{While the LB can be theoretically approached, the interplay of coherence, control complexity, and finite-time constraints fundamentally prevents its exact realization in realistic settings.}



\subsection{Non-equilibrium Landauer \rr{Erasure} }
The miniaturization in modern technology has led to the \rr{control} of small systems that are out-of-equilibrium in classical~\cite{jarzynski2011equalities,seifert2012stochastic} as well as in the quantum regime~\cite{esposito2009nonequilibrium,RevModPhys.83.771,goold2016role}. The fluctuation relation~\cite{jarzynski2011equalities,jarzynski1997nonequilibrium,jarzynski2004nonequilibrium,crooks1999entropy,tasaki2000jarzynski,kurchan2000quantum,mukamel2003quantum,campisi2011colloquium} plays \rr{the central} role in understanding the thermodynamics of these small systems that \rr{are far from the} equilibrium. Here, we mainly focus on \rr{nonequilibrium} erasure protocols within the quantum regime.

\bb{A key challenge in non-equilibrium thermodynamics is characterizing the dissipation associated with quantum erasure. Reeb and Wolf~\cite{reeb2014improved} have established a fundamental lower bound for heat dissipation, \rr{using the tools of quantum information theory and quantum statistical mechanics, that includes corrections arising from finite-size reservoir and system-bath correlation. Though Reeb and Wolf have not incorporated the fluctuations arising due to the finite-size effects in their derivation, one might be tempted to think that the correction due to the finite-size effects is somewhat related to the nonequilibrium fluctuations, as we are far away from the thermodynamic limit. Indeed, an equivalent form Eq.~\eqref{LPeq2}, has been derived independently in~\cite{Esposito_2010} to express irreversible entropy production of a finite system out of equilibrium. This indicates the necessity of LB-like bound in far from equilibrium conditions, considering the fluctuations explicitly in the setup.}  \rr{To bridge this gap, researchers have sought to establish generalized bounds for information erasure beyond equilibrium, employing the concepts of fluctuation relations~\cite{jarzynski2011equalities,jarzynski1997nonequilibrium,jarzynski2004nonequilibrium}.}} 

In~\cite{goold2015nonequilibrium}, an erasure protocol that involves a finite-size environment is \rr{studied, and a modified bound considering the fluctuations during the protocol is proposed.} This \rr{has} opened the door for the analysis of the cost of computation in non-equilibrium circumstances. Under the same four assumptions that have been used in~\cite{reeb2014improved} to derive Eq. (\ref{LPeq2}), the authors of~\cite{goold2015nonequilibrium} have first derived the following heat fluctuation-like relation of the environment for the erasure protocol
\begin{equation}\label{NONEQ1}
    \langle e^{-\beta Q} \rangle= \int e^{-\beta Q} P(Q)dQ = \sum_l \text{tr}[A^\dagger_l \rho_{\mathbf{E}}A_l] =  \text{tr} [\textbf{A} \rho_{\mathbf{E}}].
\end{equation}
 \rr{Here} $P(Q)$ denotes the heat distribution of the environment, $\bb{A_{l=jk}}= \sqrt{\lambda_j} \langle s_k |U|s_j\rangle$, with $\lambda_j$ being the eigenvalue and $|s_j\rangle$ being the eigenstate of $\rho_{\mathcal{S}}$, \rr{$U$ is the joint unitary operator acting on the system and environment} \rr{and $\textbf{A}=\sum_l A_lA^\dagger_l$}.
 The above equation is in the same spirit of  Jarzynski equality which connects the average exponentiated work to the equilibrium free energy exploiting the work distribution~\cite{jarzynski1997nonequilibrium}.
  If the operator $\textbf{A}$  is expanded in terms of the initial state of both $\mathcal{S}$ and $\mathbf{E}$ under the action of the global unitary evolution, the heat fluctuation relation becomes
 \begin{equation}\label{NONEQ2}
      \langle e^{-\beta Q} \rangle =  \text{tr}[\textbf{M} \rho_\mathcal{S}],
 \end{equation}
where $\textbf{M}=  \text{tr}_{\mathbf{E}} [U^\dagger \mathbf{I}_\mathcal{S} \otimes \rho_\mathbf{E} U$]. The desired heat dissipation during the erasure process \rr{satisfies}\footnote{ In this nonequilibrium LP section, we denote the averaged heat inside bra-ket to distinguish it from the heat exchange in a single trajectory. In Eq. (\ref{LPeq1}), Eq. (\ref{LPeq2}), and all other sections of the review, $Q$ (``heat") denotes heat dissipated on average.} 
\begin{equation}\label{NONEQ3}
    \beta \langle Q \rangle \geq \mathbf{B}_Q,
\end{equation}
where $\mathbf{B}_Q = - \,\ln (  \text{tr} [\textbf{A} \rho_{\mathbf{E}}]) = - \, \ln ( \text{tr}[\hat{\textbf{M}} \rho_\mathcal{S}])$, termed as the \textit{thermodynamic bound} in~\cite{campbell2017nonequilibrium}. The process is unital iff $\textbf{A} = \mathbf{I}_\mathbf{E}$, \rr{and therefore the bound can be insightful to compare the heat fluctuations of the environment under unital and nonunital channels}.  The operator $\textbf{A}$ is dependent on the choice of the state of the system. So, for the execution of the erasure process of a state of choice, one needs to compute $\textbf{A}$ for every instance. On the other hand, $\textbf{M}$ as defined in Eq.~\eqref{NONEQ2} can be evaluated just by implementing the unitary interaction.

\begin{figure}[h]
  \includegraphics[width=0.85\columnwidth]{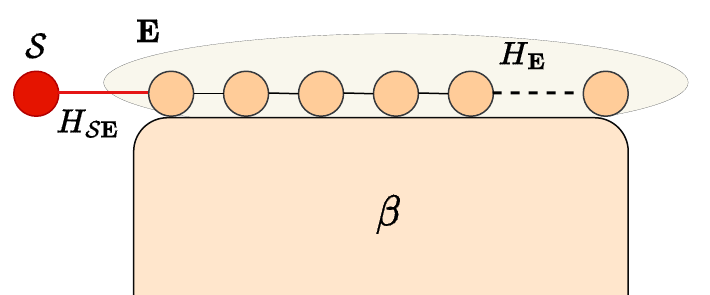}
  \caption{The schematic of the process where the system $\mathcal{S}$ interacts with the environment $\mathbf{E}$. }
  \label{Lan4}
  \end{figure}

This proposed bound is tested~\cite{goold2015nonequilibrium} on a physical system where the system is considered to be a single qubit and the environment comprises an interacting spin chain (as shown in Fig.~\ref{Lan4}). The Hamiltonian of the environment, given by the $XX$ model, is
\begin{equation}\label{NONEQ4}
    H_\mathbf{E} = J \sum_{j=1}^{N-1} \left( \sigma_x^j \sigma_x^{j+1} + \sigma_y^j \sigma_y^{j+1}\right) + B \sum_{j=1}^N \sigma_z^j.
\end{equation}
$J$ denotes the coupling strength of the interspin, $B$ is homogeneous external magnetic field and $\sigma$ describes the Pauli spin operators. The heat dissipation during the execution of the erasure protocol for this physical system is evaluated to be 
\begin{eqnarray}\label{NONEQ5} \nonumber
\beta \langle Q \rangle & = & B \sin^2 (2Jt) (2 \alpha^2 + \tanh (\beta B) - 1), \\ 
\mathbf{B}_Q & = & - \ln \left[2 (1- \phi^2) (\alpha^2 + p + 2p\alpha^2) + \phi^2 \right].
\end{eqnarray}
Here $\phi = \cos (2J t)$ \rr{at time $t$}, $p = [1 + \tanh (\beta B)]/2$ and \rr{$\alpha  (\in \mathbb{R})$ is an initial state parameter of the system}. For $\alpha \approx 1$ and for a particular temperature it is observed that $\mathbf{B}_Q \geq \mathbf{B}_{rw}$, where $\mathbf{B}_{rw}$ denotes the bound proposed in~\cite{reeb2014improved}. \rr{It is important to mention that the tightness of the bound $\mathbf{B}_Q$ over $\mathbf{B}_{rw}$ is parameter dependent. There can be regions in the parameter space where the bound $\mathbf{B}_{rw}$ is tighter than the thermodynamic one.}

\rr{In~\cite{campbell2017nonequilibrium}, the authors have performed a detailed comparative study of the thermodynamic bound $\textbf{B}_Q$ and the bound $\beta \langle Q\rangle\geq \Delta S$ (see Eq. (\ref{LPeq2})), dubbed as the entropic bound. They have explored the role of the initial state parameters on the tightness of the two bounds with respect to the dissipated heat. It has also been shown that the thermodynamic bound is not sensitive to the initial state coherence, in contrast to the entropic bound. Moreover, they have shown that there are regions where both bounds are ineffective, being less informative than the Clausius statement of the second law.}

\rr{In~\cite{taranto2018emergence}, the authors have reported a very important aspect of heat fluctuations in erasure protocol. Building upon the framework of~\cite{goold2015nonequilibrium}, they have shown using generalized concentration of measure that fluctuations of heat from its mean are exponentially damped with the size of the larger subsystem (system or environment), and this suppression scales linearly with increasing inverse temperature of the environment, regardless of the system's microscopic dynamics.}

 
\rr{Though~\cite{goold2015nonequilibrium} has paved the way to incorporate the fluctuations-like relations of heat to get an improved bound on the dissipated heat, there remains plenty of scope for further explorations. For example, the quest for a bound that considers fluctuations of heat, which is more informative than the Clausius inequality in general, remains. Generalization of these bounds to an initially correlated system-environment would be challenging and intriguing.}




The full counting statistic~\cite{esposito2009nonequilibrium} formalism \rr{has been} applied to analyze the bounds (lower and upper) on the average heat dissipation in an erasure process in~\cite{guarnieri2017full}. A single-parameter \rr{family of} bounds is developed that can be made arbitrarily tight and is independent of the \rr{underlying} map used for the execution of the process~\cite{guarnieri2017full}. This inherently marks the difference of the lower bound of the \rr{nonequilibrium erasure} process over the previous bound~\cite{goold2015nonequilibrium}. For this formalism, the minimal set of assumptions is considered as in~\cite{reeb2014improved}. 

In this full counting statistics of dissipated heat formalism, the lower bounds for the heat dissipation can be given using the (CGF)~\cite{rockafellar1970convex} as
\begin{equation}\label{NONEQ6}
    \beta \langle Q\rangle_t \geq -\frac{\beta}{\eta} \Theta(\eta,\beta,t) \equiv \mathbf{B}_Q^\eta (t) \quad (\eta>0),
\end{equation}
$\eta$ being the counting parameter. The Landauer-like bounds here are proposed with one parameter on a two-time measurement protocol. For $\eta = \beta$ the result of~\cite{goold2015nonequilibrium} is recovered.

\begin{figure}[h]
  \includegraphics[width=0.85\columnwidth]{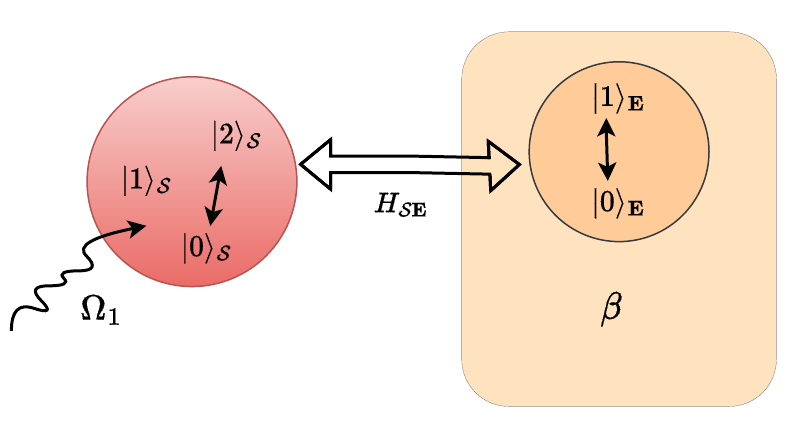}
  \caption{The schematic of the process with a three-level system and a two-level environment.}
  \label{Lan5}
  \end{figure}

Furthermore, using the large deviation function, 
which is a powerful tool to study statistical properties for a long time scale~\cite{garrahan2010thermodynamics,lesanovsky2013characterization,pigeon2015thermodynamics}, the upper bounds of heat dissipation are given as
\begin{equation}\label{NONEQ7}
    \beta \langle Q\rangle_t \leq \frac{\beta}{\vert \eta \vert} \Theta(\eta,\beta,t) \equiv \mathbf{\Tilde{B}}_Q^\eta (t) \quad (\eta <0). 
\end{equation}
 These bounds are tested on a physical system where the system is considered to be a three-level V-system and the environment is modeled by a two-level system (as in Fig.~\ref{Lan5}). The transition is pumped by a transition frequency $\Omega_1$.
This physical model highlights the tightness of the proposed bounds.

\rr{In~\cite{zhang2023quantum}, the authors have studied thermodynamics of nonequilibrium reservoirs with the help of Landauer-like bound. They have shown that a bound in terms of equality, called the Landauer-like bound, always holds even for nonequilibrium reservoirs, but an inequality obtained from the equality by excluding a few contributions might be violated. Violation of this inequality, dubbed as Landauer-like inequality, indicates the possibility of spontaneous heat transfer from a cold bath to a hot bath, beating the Carnot bound and extracting work from a single nonequilibrium reservoir.}

\subsection{Landauer Bound under Non-Markovian Dynamics}

The Markovian approximation is the most convenient way to express the dynamics of OQS~\cite{rau1963relaxation,alicki2007quantum,scarani2002thermalizing,ziman2005all,gennaro2009relaxation}. In this process, the evolution timescale \rr{of the system} is considered to be larger than the correlation time of the environment. In other words, in this approximation, the memory effect (or information backflow) is neglected. However, it plays a non-minuscule role in the analysis of the dynamics of the system and thermodynamic cost of \rr{erasure}. \rr{This backflow of information or non-Markovian (NM)~\cite{de2017dynamics,breuer2016colloquium} dynamics has been studied following various methods and approaches~\cite{breuer2009measure,rivas2010entanglement,chruscinski2014degree}.} 

 \bb{Under NM dynamics, where coherence endures and information retrogrades from the environment, a coherence-dependent correction to LB~\cite{reeb2014improved} becomes imperative. Unlike Markovian erasure, where dissipation follows a strict lower bound, NM effects can modify these limits, potentially reducing or amplifying heat dissipation.}
\bb{Here, we discuss LP \rr{under} NM dynamics, exploring how memory effects influence erasure costs and coherence-driven corrections. This provides a more comprehensive view of quantum erasure beyond the standard Markovian framework.}

Out of the various methods for \rr{exploring NM dynamics}, one of the \rr{prominent approaches is adoption of the} collision model for OQS~\cite{rau1963relaxation,alicki2007quantum,scarani2002thermalizing,ziman2005all,gennaro2009relaxation}. The process of analyzing the information to energy conversion in the collision model provides the platform for the foundation of the LP under NM dynamics.
The formulation of a Ladauer-like principle for heat flux in an erasure process delineated by the collision model \rr{has been} first proposed in~\cite{lorenzo2015landauer}. 
 \begin{figure}[h]
\includegraphics[width=1.0\columnwidth]{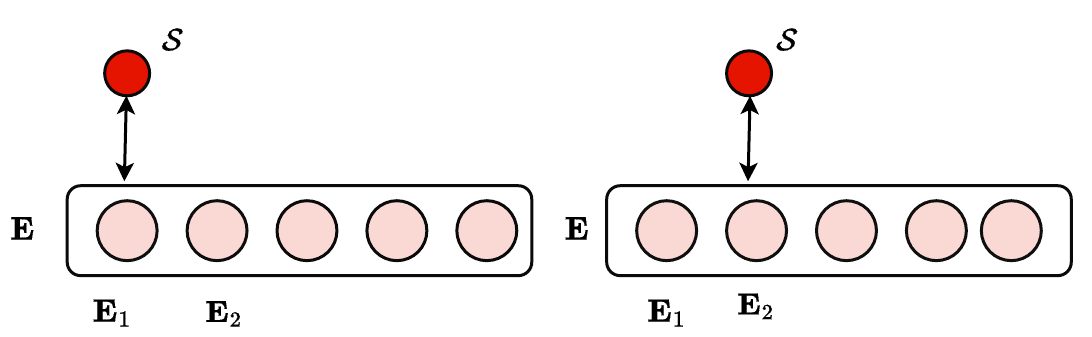}
\caption{The system $\mathcal{S}$ interacts with the particle of the environment $\mathbf{E}$ which is termed as the subenvironment. Then the system undergoes pairwise collision with the subenvironment with the restriction that it will not interact with an ancilla of the environment twice. }
  \label{Lan1}
\end{figure} 
For the execution of the protocol~\cite{lorenzo2015landauer}, the thermalization of the system $\mathcal{S}$ with the environment $\mathbf{E}$ is considered (as shown in Fig.~\ref{Lan1}). In this model, the environment has N identical noninteracting elements $\mathbf{E}_N$, which are conveyed as \textit{subenvironments} (Fig.~\ref{Lan1}), and each of them is considered to be in a thermal state $\eta = \otimes_{n=1}^N \eta_n^{th}$, where $\eta_n^{th} = e^{-\beta H_{E_n}}/Tr[e^{-\beta H_{E_n}}]$. Here, $H_{E_n}$ denotes the free Hamiltonian of the n-th \textit{subenvironment} and $H_\mathcal{S}$ denotes the free Hamiltonian of the system. The system interact with the environment, and the interaction \rr{is governed by} a sequence of pairwise collisions with the \textit{subenvironments}. The environment is of a large size to curtail the situation that the system interacts with the same \textit{subenvironment} twice. Each collision of this process is described by unitary evolution $U = e^{-igH_{int}\tau}$, where $g$ is the interaction strength, $\tau$ is the interaction time, and $H_{int}$ denotes the \rr{time-independent} interaction Hamiltonian of the system and the environment. The information that is stored in the system gets diluted while interacting with the environment. After (n+1) collisions, the state of the system and the environment are respectively described as
\begin{eqnarray} \label{NONM1} \nonumber
\rho_{n+1} = \text{Tr}_{E} [U \rho_n \otimes \eta_{n+1}^{th} U^\dagger] = \Phi [\rho_n], \\ 
\eta_{n+1} = \text{Tr}_{S} [U \rho_n \otimes \eta_{n+1}^{th} U^\dagger] = \Lambda_n [\eta_{n+1}^{th}],
\end{eqnarray}
where $\Phi$ and $\Lambda_n$ denote the completely positive trace-preserving (CPTP) map. The variation of energy of the system and the heat exchange with the environment are
\begin{eqnarray}\label{NONM2} \nonumber
\Delta \mathcal{E}_{n+1} =\text{Tr} [H_{\mathcal{S}} (\Phi - \mathbf{I})[\rho_n]], \\
\delta Q_{n+1} = \text{Tr} [H_\mathbf{E} (\Lambda_n - \mathbf{I})[\eta^{th}]].
\end{eqnarray}
Now, if energy-conserving interaction between the system and the environment is considered and $[U, (H_{\mathcal{S}} + H_\mathbf{E})]=0$ is assumed, then we have $\dot{Q} = -\dot{\mathcal{E}}$, \rr{where the $``."$ denotes the rate of change with time}. The stationary state of the system is described by the Gibbs state $\rho^{\text{eqb}} =  e^{-\beta H_{\mathcal{S}}}/Tr[e^{-\beta H_{\mathcal{S}}}]$ with the initial temperature of the bath. The \rr{quantum} relative entropy $S(\rho||\rho^{\text{eqb}})$ among the state at a time $t$ with the stationary state obeys  $S(\rho||\rho^{\text{eqb}}) = - \dot{S} (\rho) + \beta \mathcal{\dot{E}}$. As the relative entropy is non-increasing under the CPTP map~\cite{vedral2002role}, one obtains
\begin{equation} \label{NONM3}
    \beta \dot{Q}(t)\geq \dot{\Tilde S} (\rho),
\end{equation}
where $\Tilde S (\rho) = - S (\rho)$. Eq.~\eqref{NONM3} provides the \rr{LP in OQS formulation involving heat and entropy fluxes}. \rr{This relation} provides a time-resolved analysis of the erasure by \textit{thermalization}. \rr{Moreover}, it is possible to elucidate the role of correlation \rr{among the subsystems, contributing through the entropy,} in the information-erasure processes. \rr{This in turn enable us to probe the effect of the NM dynamics in information-erasure.}


The \rr{local} violation of the LB due to \rr{the system-environment} correlation (proposed in~\cite{lorenzo2015landauer}) is \rr{demonstrated} in the spin-1/2 particle system~\cite{pezzutto2016implications} when \rr{interaction between the system and environment is governed by the Heisenberg interaction}. 
The sequence of discrete-time collisions of the system with one of the environmental particles at a time occurs in this process. In the long time limit, the environment (for non-interacting particles) undergoes \textit{homogenization dynamics}, \rr{i.e., in the long time limit which is equivalent to a large number of collisions of the system with the environment, the state of the system approaches asymptotically to the initial preparation of the environment}~\cite{ziman2001quantum,scarani2002thermalizing}. If the environment is considered to be composed of interacting particles, the system undergoes an NM \rr{dynamics}. The elements of the environment are in a thermal state, and in the asymptotic limit, \textit{homogenization dynamics} are encountered. This behavior occurs when the state of the elements of the environment has weak fluctuations. However, due to the inter-environment interactions, there is a memory effect in the dynamics of the system. 
It \rr{has been} observed that \rr{during the time evolution, the system-environment correlation grows strong enough to exhibit NM features leading to} an instantaneous violation of the LB.

\begin{figure}[h]
  \includegraphics[width=0.8\columnwidth]{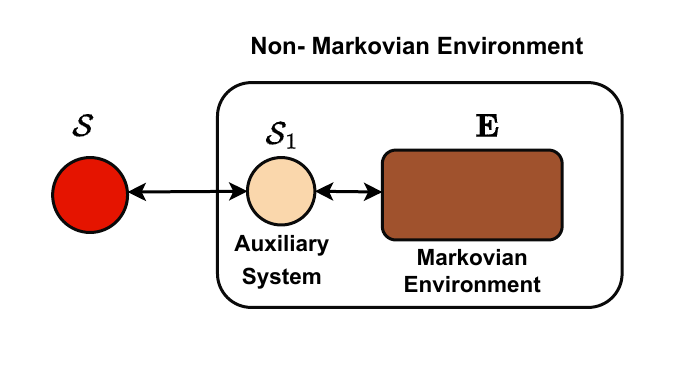}
  \caption{The system $\mathcal{S}$ interacts with the auxiliary system  $\mathcal{S}_1$, which then interacts with the environment $\mathbf{E}$ and is redirected to its initial value after each collision. The auxiliary system and the environment form the non-Markovian environment.}
  \label{Lan2}
  \end{figure} 

The explicit analysis of the cause and the condition for the violation of the LP in the NM environment is explicated in~\cite{man2019validity}. 
A modified form of the collision model is considered in~\cite{man2019validity}, where the system ($\mathcal{S}$) information is first transferred to the subenvironment via the system-subenvironment collision. A part of this information gets transferred to the next subenvironment via intra-collision. This intra-collision leads to the system-environment correlation, and then the system interacts with the next subenvironment. So, there is prior information of the system in the subenvironment before the next system-subenvironment collision. This causes the non-Markovianity in the process. The system-environment collision is governed by a unitary operation $U$. The change in entropy of the system due to the interaction with the subenvironment is
\begin{eqnarray}\label{NONM4} \nonumber
\Delta S_n & = & S(\rho_{\mathcal{S}\mathbf{E_n}}^\prime ||\rho_{\mathcal{S}_n}^\prime \rho_{\mathbf{E_n}}^{th})+ \text{Tr}_{E_n} [(\rho_{\mathbf{E_n}}^{\prime}- \rho_{\mathbf{E_n}}^{th})\,\, \ln \,\rho_{\mathbf{E_n}}^{th}] \\
& - & I(\rho_{\mathcal{S} \mathbf{E_n}}) - S (\rho_{\mathbf{E_n}} || \rho_{\mathbf{E_n}}^{th})),
\end{eqnarray}
where $S (\rho || \rho^\prime) = \text{Tr}\, [\rho \,\ln \, \rho] - \text{Tr}\, [\rho \,\ln \, \rho^\prime]$ describes the quantum relative entropy, $\rho_{\mathcal{S}_n}^\prime$ and $\rho_{\mathbf{E_n}}^{\prime}$ denotes the marginal states of the system and the environment respectively after the collision. $I(\rho_{\mathcal{S} \mathbf{E_n}})$ characterizes the mutual information that quantifies the system-environment correlation based on the intra-collision strength. The second term in Eq.~\eqref{NONM4} describes the entropy flow from the environment to the system. 
The \rr{modified} LB  for the NM process \rr{can be expressed as}
\begin{eqnarray}\label{NONM5} \nonumber
 \beta_{\mathbf{E_n}} \Delta \Tilde{Q_n} & = & \Delta \Tilde{S}_n + S(\rho_{\mathcal{S}\mathbf{E_n}}^\prime ||\,\rho_{\mathcal{S}_n}^\prime \rho_{\mathbf{E_n}}^{th}) 
-  I(\rho_{\mathcal{S} \mathbf{E_n}})\\ & - & S (\rho_{\mathbf{E_n}} ||\, \rho_{\mathbf{E_n}}^{th})),
\end{eqnarray}
where $\Delta \Tilde{Q_n} = -\Delta Q_n$ \rr{is the heat dissipated to the environment} and $\Delta \Tilde{S}_n = - \Delta S_n$ \rr{entropy decrease of the system}. When $I(\rho_{\mathcal{S} \mathbf{E_n}}) \leq S(\rho_{\mathcal{S}\mathbf{E_n}}^\prime ||\rho_{\mathcal{S}_n}^\prime \rho_{\mathbf{E_n}}^{th}) - S (\rho_{\mathbf{E_n}} || \rho_{\mathbf{E_n}}^{th}))$, the LP holds, \rr{or we can say that} the established system-environment correlations are smaller than the upper bound.  \rr{Therefore}, the condition that allows 
to violate the LP is  
\begin{equation} \label{NONM6}
    I(\rho_{\mathcal{S} \mathbf{E_n}}) > S(\rho_{\mathcal{S}\mathbf{E_n}}^\prime ||\,\rho_{\mathcal{S}_n}^\prime \rho_{\mathbf{E_n}}^{th}) - S (\rho_{\mathbf{E_n}} ||\, \rho_{\mathbf{E_n}}^{th}).
\end{equation}
\rr{The authors have demonstrated violation of LP by taking the system-environment interaction to be the Heisenberg interaction. Their findings further suggest that the complexity of the environment has little to do with the violation of LP under NM dynamics.} \rr{They have further shown that their results} can be generalized \rr{to} where the system is coupled to multiple environments. The LP violation condition for the multiple NM environments is found to be \rr{a straightforward generalization of} Eq.~\eqref{NONM6}.

A solution to the \rr{apparent} violation of the LP \rr{in the NM domain} has been put forward in the work~\cite{zhang2021non}. Here, the authors considered a different scenario, the system ($\mathcal{S}$) interacts with $\mathcal{S}_1$ (ancillary system) which in turn is coupled to a Markovian environment  $\mathbf{E}$. This composite environment induces NM dynamics (Fig.~\ref{Lan2}) \rr{through the memory effect induced by $\mathcal{S}_1$.}
 In the Markovian limit, the conventional LB holds. The modified version of the LB for the heat dissipation that holds for the NM as well as the Markovian regime is
\begin{equation} \label{NONM7}
    \beta \dot{Q}_\mathcal{S} (t) \geq \dot{\Tilde{S}} (\rho_{\mathcal{S}} (t)) + \dot{I} (\rho_{\mathcal{S^\prime}} (t)) + \dot{S} (\rho_{\mathcal{S}_1} (t) || \rho_{\mathcal{S}_1}^{th}),
\end{equation} 
where $\dot{Q}_\mathcal{S} (t)$ describes the heat flux from the system to the environment, $\dot{\Tilde{S}} (\rho_{\mathcal{S}}(t))$ is the entropy flux from $\mathcal{S}$ to
$\mathbf{E}$ ($\Tilde{S} (\rho_{\mathcal{S}}(t)) = - S(\rho_{\mathcal{S}}(t))$), $\dot{I} (\rho_{\mathcal{S^\prime}} (t))$ describes the rate of mutual information that characterizes the correlation between the system and auxiliary system and $\dot{S} (\rho_{\mathcal{S}_1} (t) || \rho_{\mathcal{S}_1}^{th})$ describes the rate of quantum relative entropy. \rr{Their model allows them to study the Markovian limit, and it has been shown that in the Markovian limit the above equation reduces to \eqref{NONM3}.} The generalization of the model for multiple environments has also been reported in this work. The modified LP for the multiple environment case is also found to be valid in both regimes. \rr{Importantly, the quantum information theoretic framework that has been adopted to derive the the modified LBs in the NM domain discussed here broadly belongs to the same framework used to derive Eq.~\eqref{LPeq2}}.

 In~\cite{hu2022relation}, the authors \rr{have} evaluated the connection between the information back-flow in the NM dynamics and \rr{the violation of} LP. For the analysis, a qubit is considered to be coupled with the environment. If the system is initially in the thermal state, it is inferred from the analysis that there is a one-to-one correspondence between the violation of the LP and the information backflow. Whereas, the correspondence does not hold if the initial state of the system has coherence. 

\section{Landauer limit in Computing}\label{sec4}
A quantum computer~\cite{nielsen2002quantum,aharonov1999quantum,divincenzo2000physical} harnesses quantum mechanical \rr{resources} such as superposition and entanglement~\cite{mermin1990extreme1,linden2006entanglement,wineland2013nobel1} to perform computations. Qubits serve as the fundamental units of information in quantum computers and leverage the principles of quantum superposition enabling quantum computers to execute certain types of calculations much more efficiently than classical computers. However, regardless of the computational task, it ultimately needs to be implemented on a physical system, implying that computations are subject to the constraints of the laws of physics. 
For instance, the LP places a minimum limit on the heat generated during \rr{a} bit erasure, and the \textit{quantum speed limit}~\cite{golub2014scientific,caneva2009optimal,okuyama2018quantum,jones2010geometric,deffner2017quantum,del2013quantum,deffner2013quantum} dictates a limitation on how quickly a fundamental logical operation can be carried out. 


\subsection{Landauer Limit For N-based logical computer}
A computer using the binary \rr{logic} can be exemplified as a single particle Szilard engine~\cite{szilard1929entropieverminderung} where the certainty of the particle's presence in a chamber \rr{among the two} corresponds to the recording of 1 bit of information, and the uncertainty in the particle's location corresponds to the erasure of 1 bit of information. However, computers are not limited to binary logic systems. In principle, they can be based on many-valued logic. For instance, a ternary logic-based computer (``\textit{trit}") is founded on the concept of a ternary symmetrical number system and ternary memory element (``flip-flap-flop")~\cite{glusker2005ternary,brousentsov1965experience,stakhov2002brousentsov,frieder1973balancedternary,knuth1998art}. Recently, computers employing many-valued logic have garnered significance due to \rr{both} fundamental aspects and numerous applications~\cite{gottwald2001treatise,chang1958algebraic}. 

\begin{figure}[h]
  \includegraphics[width=0.65\columnwidth,height=0.5\columnwidth]{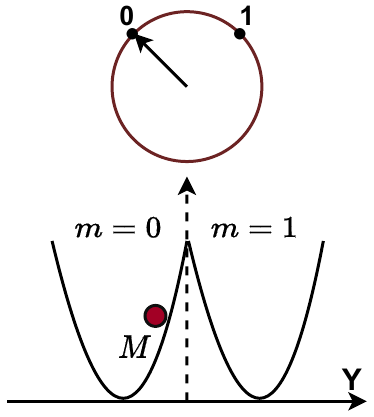}
  \caption{Schematic of the qubit model with a Brownian particle $M$ placed in a double potential well. The qubit can be trapped in either of the sides of the well. The corresponding information states are $m=0$ and $m=1$.}
  \label{Landauer_erasure_3}
  \end{figure}

The exemplification of the LP for binary logic computers can be supplied by the Brownian particle in a double potential well as shown in Fig.~\ref{Landauer_erasure_3}.  For a symmetrical well and a \rr{completely} random bit with $a_0 = a_1 =\frac{1}{2}$, \rr{where $a_0$, $a_1$ denotes the probability of being trapped in the wells,} the LB is reached. Now, whether \rr{LB} holds for N-based logic is explored in~\cite{bormashenko2019generalization}, \rr{where} the authors have considered ternary logic, i.e., \textit{trit} computing element for the analysis which is further generalized for N-based logic. Similar to binary logic computers, the LP is illustrated for ternary logic computers using a Brownian particle in a symmetric triple-well potential. For a random bit, $a_0= a_1= a_2= \frac{1}{3}$, \bb{where $a_0$, $a_1$, $a_2$ denotes the probability \rr{of being trapped in the wells}}, the LB is reached for the ternary logic, and this result has been further extended to N-based logic.

As computational algorithms become more complex, parallel computing becomes essential for the efficient execution of protocols. Parallel computing, as discussed in~\cite{kumar1994introduction,barney2010introduction,melhem1992introduction,golub2014scientific}, involves the simultaneous utilization of multiple processors to solve algorithms. The primary objective is to distribute the workload among several processors to solve problems more swiftly or handle larger problems within the same time window. In~\cite{konopik2021fundamental}, the authors examined the energy cost of finite-time computing using non-equilibrium thermodynamics principles. It has \rr{been} shown that the energy cost for computational tasks in a parallel computer, within a given finite time, closely adheres to the \rr{LB}. This cost remains bounded even as the computational problem size increases.



\subsection{Landauer \rr{Limit} in Presence of Time-symmetric Protocol}

Investigations into energy dissipation with time-symmetric protocols, \rr{demonstrates a fundamental trade-off between computational accuracy and energy cost~\cite{riechers2020balancing}. It shows that reducing logical error requires increasing energy dissipation, particularly under nonreciprocal operations, and therefore} time-symmetric \rr{protocols} prevent computation from reaching the LB. In a subsequent work~\cite{wimsatt2021refining}, authors \rr{did} thermodynamic analysis of time-symmetric procedures (as prescribed in~\cite{riechers2020balancing}) to thoroughly examine the trade-offs between accuracy and dissipation during information erasure. The authors employed nonequilibrium information thermodynamics to calculate the minimum energy dissipation needed for reliable erasure under time-symmetric control protocols. The energy costs associated with \rr{this} reliable erasure were found to be higher than those implied by the LB. Moreover, these costs diverge in the limit of perfect \rr{erasing}. Therefore, the creation of time-asymmetric protocols is deemed necessary for effective and precise thermodynamic \rr{analysis of} computation. Consequently, it is concluded that time asymmetry serves as a crucial design principle for thermodynamically efficient computing, warranting \textit{further investigation}.



\section{Experimental Validation of Landauer's Principle}\label{sec.5}
\rr{With our ever-increasing control over the miniaturized scale} and  
the advancement of quantum technology, \rr{the validation of the LB is becoming more and more plausible day-by-day}. Here in this section, we will focus on the recent experimental advancement to test LP in different technological platforms.

\subsection{Optics \rr{Based} Technology}

\textbf{Optical Tweezers:}

\begin{figure}[h]
  \includegraphics[width=0.8\columnwidth,height=0.4\columnwidth]{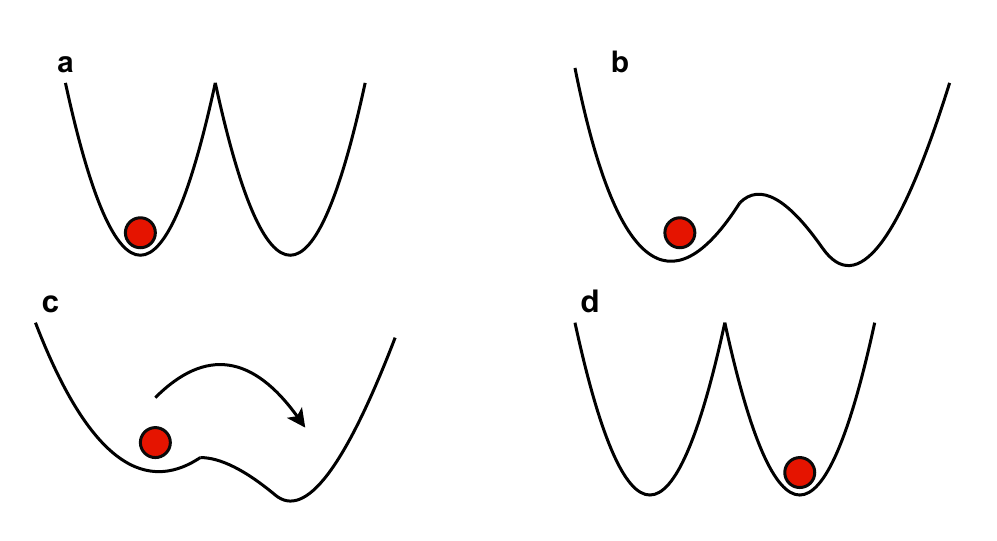}
  \caption{The erasure protocol that is considered in the experiment. }
  \label{Lan3}
  \end{figure}

The test of LP \rr{has become} possible (that remained untested over five decades) after the two basic advancements. One of them \rr{is} to develop a method to investigate the work done on the particle as well as the heat dissipation by the particle based on the information on the trajectory of the particle and its potential. It \rr{has been} proposed and tested in the seminal paper by Sekimoto~\cite{sekimoto1997kinetic,sekimoto2010stochastic}. The second advancement is the development of methods to impose user-defined potential on small particles. For e.g., the usage of the localized potential force. This potential is created by optical tweezers and is used to explain LP under partial erasure~\cite{berut2012experimental}. In this work, the authors have considered an overdamped colloidal particle inside the double potential well (Fig.~\ref{Lan3}) which is created by focusing a laser alternatively at two different positions with a high switching rate. 
The form of the potential well is determined by the intensity of the laser and the distance between the two focal points. If the particle is on the left side of the well, the state of the system is denoted by `0', whereas if the particle is on the right side, it is `1'. The experimental process for this method can be summarized as follows:

$\bullet$ Initially the \textit{bead} is considered to be trapped in one of the wells with a definite state. The central barrier is kept high so that the jumping time is very high.

$\bullet$ Now the intensity of the laser is reduced so that the barrier is low enough and the \textit{bead} can jump from one to another. 

$\bullet$ Finally, after the \textit{bead} ends in the required well independent of the initial state (this causes the memory erasure when it is set to 1), the barrier is raised to its previous stage. 

Following the same methodology, the authors in their work~\cite{jun2014high} have embraced a more flexible approach with a feedback loop to create virtual potential. An anti-Brownian electrokinetic (ABEL) feedback trap is used for testing LP. This model provides the advantage of measuring the work with high precision for testing LP.  Berut et. al.~\cite{berut2012experimental} in their work were not able to acquire the complete erasure as they did not have full control over the potential shape. It was reported that in the asymptotic limit, LB is $\approx (0.13- 0.49) k_B T$ whereas the full-erasure limit is $0.69 k_B T$. 
Jun et. al.~\cite{jun2014high}, in their work, reports the complete erasure and shows that their approach results in \rr{reaching} LB. A complete and detailed analysis of the various contribution that causes heat dissipation in the system is reported~\cite{berut2015information} in verifying LP. 

In~\cite{PhysRevLett.117.200601}, the authors have explored the erasure principle for the symmetry-breaking case by analyzing an asymmetric double-well potential. The analysis of this process, following the methodology of Sagawa and Ueda~\cite{sagawa2014thermodynamic}, conveyed that to erase a bit of information the average work can be less than $k_B T \ln2$ provided that the volume of the phase space for each state is different. The memory cell in the experiment consists of an overdamped silica bead trapped in the double-well potential foist by ABEL trap. It is encountered in the experiment that the work for this asymmetric bit erasure can be less than that of $k_B T \ln2$. 

\textbf{Interferometer:}

To explore the LP for the underdamped and overdamped systems~\cite{dago2021information,dago2022dynamics}, \textit{differential interferometer}~\cite{paolino2013quadrature} has been considered as the platform for the analysis.  The working system is a \textit{micromechanical oscillator} (the role played by a conductive cantilever in the experiment) confined in a double potential well.  Here, the cantilever is the memory cell. Within the framework of stochastic thermodynamics, it has been shown in the work~\cite{dago2021information} that one can reach the LB with high precision within a time scale of 100 ms, which was previously recorded to be 30 s. The work was further extended by the authors to explore the overdamped and underdamped case within the stochastic thermodynamic framework with fast operation~\cite{dago2022dynamics}. There they encountered a transient temperature rise, so the mean work to erase the information increases but is still bounded by LP. 

\rr{\textbf{Magneto-Optical Kerr effect:}}

\textit{Nano-magnetic switches} are one of the prime components that will play an extensive role in electronic applications like storage media. The information in such devices is encoded by electron spin. It is a bistable switch that comprises elongated ferromagnetic dots of various sizes and shapes. The authors in~\cite{martini2016experimental} have investigated the energy cost, i.e., the LL of resetting the magnetic binary switches composed of elliptical or rectangular ferromagnetic dots of different sizes and shapes. The dissipated energy in this process is measured by the vectorial \rr{magneto}-optical Kerr effect (MOKE) experiment in Permalloy (Ni$_{80}$ Fe$_{20}$)~\cite{martini2016experimental}. The logic states `0' and `1' are described by the two orientations of magnetization.  The experimental process for this method can be summarized as follows:
(a) The process starts with the equilibrium magnetization. The system is in either of the states, and
(b) several magnetic fields are applied to reduce the effect of the barrier as well as help the system jump from one to another to reach the final logic state. 

In this model, the authors encountered that there is a deviation from the theoretical limit of up to three orders for magnetic dots with dimensions of several hundred nanometers. The \textit{morphological imperfection} as well as the \textit{inhomogeneity of the magnetization} are the primary causes of the deviation.   Whereas if one reduces the dot size, it is shown to approach the theoretical LL.

Following the same technology, the authors in~\cite{hong2016experimental} have explored the intrinsic energy dissipation for a single-bit operation. They have used a nano-scale digital magnetic memory as their working system. The MOKE experimental setup is considered for the analysis of the energy dissipation during the execution of the process. \rr{In this process} the nanomagnet plays the role of the memory bit, and magnetic anisotropy is utilized to create the \textit{easy axis} \rr{along which the net magnetization aligns to minimize magnetostatic energy}. 

So far the experimental validation has been restricted in the classical domain. The LB in the quantum regime was experimentally analyzed in the work~\cite{gaudenzi2018quantum}, where a crystal of molecular nanomagnet as the spin memory is considered for the analysis. In this model, the crystal of Fe$_8$ molecular magnet~\cite{gatteschi2006molecular} plays the role of quantum spin memory which measures the energy dissipation during the execution of the erasure process. The process is equivalent to the method used in~\cite{martini2016experimental} to develop the double potential well and reset the system to the final state `1' by applying a magnetic field. The erasure of the memory is still governed by the LP. Surprisingly, it was encountered that maximum energy efficiency is achieved while preserving the fast operation process unlikely to the classical system.




\subsection{Trapped Ions}
In a quantum regime where the information is encoded in a qubit, one needs to reconstruct the model that is considered in the classical regime to be applicable in the quantum realm. The work~\cite{yan2018single} has explored the quantum LP based on trapped ion $Ca^{+}$. The ion is trapped in a linear Paul trap. Here, the LB is evaluated by the analysis of the system-reservoir correlation and the change in entropy during the execution of the erasure process. Trapped ions are considered \rr{as} one of the perfect platforms for the exploration of 
quantum thermodynamics with high accuracy~\cite{an2015experimental,huber2008employing,rossnagel2014nanoscale}. The two internal levels of the system ion are considered as the qubit system, and the vibrational modes of the ion as the finite temperature bath. The LP is analyzed in the system by observing the phonon number in the variation of the ion. The authors have confirmed that the LP holds in the quantum regime experimentally.

\subsection{Nuclear Magnetic Resonance Technology}
The process to measure the heat dissipation in quantum logic gates using the nuclear magnetic resonance (NMR) setup is proposed in~\cite{peterson2016experimental}. For the analysis, a three-qubit system is considered (the working system, environment, and the ancilla) to evaluate the heat dissipation of the process. In the first step of the process, the \textit{interferometric technique} is considered for the reconstruction of the dissipated heat using the ancilla. In the second phase of the process, the change in the entropy of the system is measured through \textit{quantum state tomography}~\cite{cramer2010efficient,christandl2012reliable,lvovsky2009continuous,gross2010quantum,stricker2022experimental,liu2012experimental}.  The system is developed by dissolving trifluoroiodoethylene (C$_2$F$_3$I) molecule in D$_6$ (97\%). The three $^{19}F$ nuclear spin forms the three-qubit system for the analysis. The extracted average heat during the execution of the process is found to be bounded by the LP.

\subsection{Superconducting Technology}

An \rr{experiment has been performed} on a hardware platform, namely, superconducting flux logic, for analyzing the quantum LP~\cite{saira2020nonequilibrium}. The double well potential in this process arises due to the \textit{Josephson effect} and the \textit{flux quantization}. The bit erasure process is explored in the parametric regime where the approximation of the metastable state is valid. It is \rr{observed} that the process is bounded by LB.

\section{Reversible Computation Model and Thermodynamic Interpretation}\label{sec.6}

\bb{In this section, we first provide a brief overview of the reversible model of computation~\cite{keyes1970minimal,likharev1982classical,bennett1982thermodynamics,bennett1989time,landauer1961irreversibility,feynman2018simulating,richard1986quantum,zurek1989thermodynamic,raussendorf2001one,bennett1985fundamental}: Ballistic computer and Brownian computer mainly, followed by their thermodynamic interpretation. We first discuss the Ballistic computer (BLC) proposed by Fredkin and Toffoli~\cite{fredkin1982conservative} and its limitations. Subsequently, we discuss the Brownian computer (BWC), which utilizes thermal fluctuation to perform a computational process.}

\rr{Before delving into reversible computation and its thermodynamic interpretation, it is worthwhile to briefly overview the two key concepts: \textit{thermodynamic reversibility} and \textit{logical reversibility}.} Thermodynamic reversibility~\cite{jarzynski1997nonequilibrium,crooks1998nonequilibrium,crooks1999entropy,jarzynski2000hamiltonian,seifert2005entropy,kawai2007dissipation,crooks2011thermodynamic,wolpert2024stochastic,sagawa2018second} can be defined as follows: A physical process is considered thermodynamically reversible if and only if the time evolution of the probability distribution in the process can be reversed. This reversal should include the time reversal of changes in external parameters, along with the inversion of the signs of both work and heat quantities. On the other hand, logical reversibility~\cite{landauer1996minimal,bennett1973logical,bennett2003notes,sagawa2014thermodynamic} is defined as follows: A computational process is logically reversible if and only if it is an \rr{bijection}. In other words, for any output logical state, there is a unique input logical state. Otherwise, it is considered logically irreversible. \rr{ According to Landauer~\cite{landauer1961irreversibility},} a positive amount of heat emission is inevitable while a logically irreversible process occurs \rr{where information is erased or thrown away}. These two fundamental concepts are crucial in the analysis of thermodynamic computation processes, underscore the pivotal role of thermodynamics in computational theory. \rr{How thermodynamic reversibility influences the heat generation in an logical irreversible process, for example the information erasure is summarized in Table~\ref{Tablew1}.}
\begin{table}[h]
\begin{center}
\begin{tabular}{ c|c|c } 
\hline
 & Quasi-static  & Finite velocity \\
\hline
Thermodynamically & Reversible  & Irreversible \\  
Heat emission & $=1/\beta \ln 2$ & $> 1/\beta \ln 2$ \\ 
\hline
\end{tabular}
\end{center}
\caption{Role of thermodynamic reversibility in logical irreversible processes.}
\label{Tablew1}
\end{table}




 
\subsection{Ballistic Computer}

\begin{figure}[h]
  \includegraphics[width=0.9\columnwidth,height = 0.7\columnwidth]{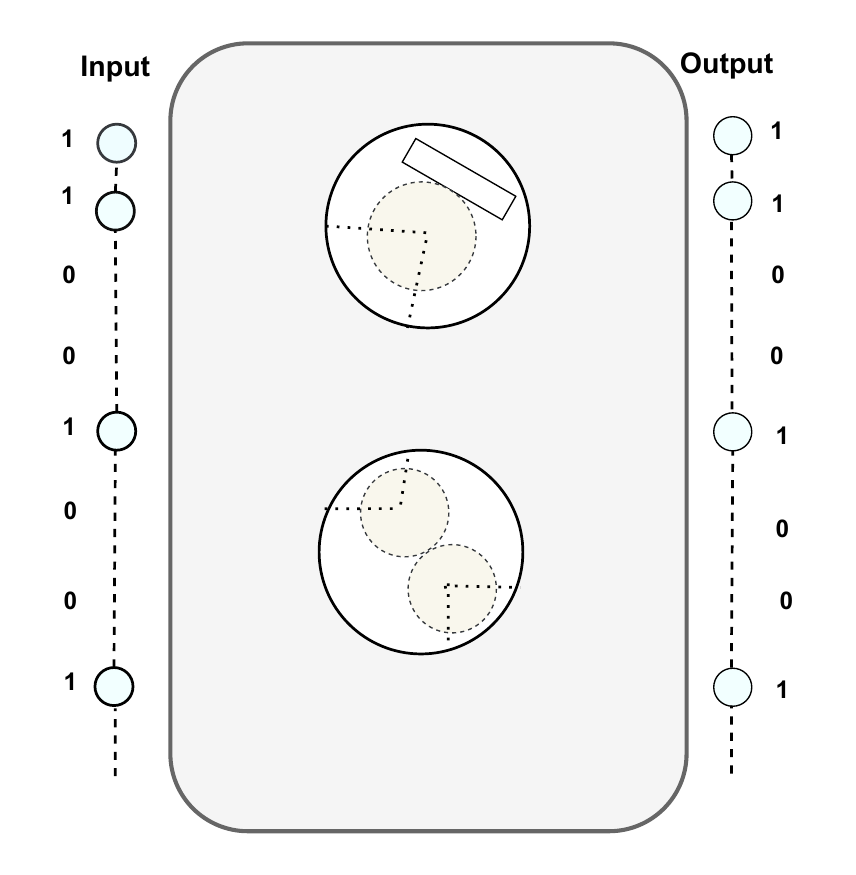}
  \caption{A schematic of a BLC proposed by Toffoli. The key condition for this model is that the number of 1's in the input must be equal to the number of 1's in the output, ensuring that the \textit{Boolean function} is both conservative and reversible in nature.}
  \label{fig5}
  \end{figure}

The principle of the \textit{ballistic} computation model~\cite{fredkin1982conservative} is based on elastic collisions. This model consists of hard spheres that collide between themselves elastically and with fixed reflective barriers. From the input side of the model, as shown in Fig.~\ref{fig5} (\textit{starting line} here), a huge number of hard spheres (balls) are fired with equal velocity. If `1' is there in the input, a ball is considered in the starting line, else no ball for `0'. Due to the collision process inside, the ball changes its direction and collides with the other balls. 
The balls, after a finite number of collisions, reach their finishing point. This signifies the output of the computer. The mirror of this computer is equivalent to the logic gates of our digital computers, and the balls are equivalent to the signals. 


Any bijective function is computable in this model, but it will be unable to compute non-conservative (non-bijective) Boolean functions~\cite{ogaard2021boolean}. Though it conveys to decrease in the amount of cost in energy, we encounter some drawbacks of this setup. The two main drawbacks of this system are its sensitivity to small perturbative changes, and the second one is related to the collision of the balls. It is quite difficult to make each collision elastic. \rr{Furthermore, these collisions result in the thermal randomness in the system.}


To address the collision problem, one approach is to correct the instability in the velocity and position of the ball after each collision process. While this provides a solution, it renders the system thermodynamically irreversible. Another method to mitigate this effect is to use square balls instead of spherical ones. This approach eliminates the exponential growth of errors as the square balls remain parallel to the wall and each other. However, it is worth noting that the use of square balls is unnatural due to the non-existence of square atoms in nature.

Quantum effects can stabilize the system from this problem, but it will bring some new instability~\cite{benioff1982quantum}. The wave-packet spreading causes instability in the system in the quantum realm. Benioff~\cite{benioff1982quantum} in his work has discussed a quantum version of the BLC, where he has proposed a way to mitigate the effect of the noise due to the wave packet spreading by utilizing a time-independent Hamiltonian.


\subsection{Brownian Computer}
As \textit{thermal randomness} is inevitable, the strategy of Brownian computers~\cite{bennett1982thermodynamics} is to exploit it. In this model, the trajectory of the dynamical part of the system is influenced by \textit{thermal randomization} in such a way that it attains Maxwell velocity \rr{distribution} and is equivalent to a random walk. Despite its chaotic nature, the BWC is able to execute valuable computations. 

The state transition for the BWC happens due to the random thermal movement of the part that carries the information. Due to its random nature, the transition can backtrace (move backward) in the computational process, undoing the transition executed recently, \rr{albeit the transition is slightly biased towards the forward direction}. In the macro regime, the execution of computation using a BWC seems counterintuitive, but this is an obvious situation in the micro regime. 

Bennett has proposed \cite{bennett1982thermodynamics} that one can execute a Turing machine (TM \rr{(see Sec.~\ref{sec7} for TM)}) using this thermal randomness. It is made up of clockwork, which is frictionless and rigid in its form. The parts of the clockwork TM should be interlocked so that they have the freedom to jiggle around locally, but are restricted from moving an appreciable amount for the execution of a logical transition. Bennett presumed that a driving force (some energy gradient) to execute the computation \rr{in less time}  and  \rr{a trap for the stability of halting state are required, and they can be done with arbitrarily small entropy generation which maintains the effective reversibility of the model}. \rr{However,} in ~\cite{norton2013brownian}, the authors argued that it has an entropic cost which renders the model irreversible.
In the context of \textit{computational complexity}, a comparable model of BWC was analyzed by  Reif~\cite{reif1979complexity} to explore the relationship between P and PSPACE (P = PSPACE )~\cite{arora2009computational}.

\subsection{Brownian Computer: Thermodynamic Interpretation}\label{sec20}

The thermodynamic analysis of the Brownian motion of particles, which are integral to the BWC, has been approached through various processes~\cite{norton2013brownian,nicolis2017stochastic,pal2020stochastic,meerson2022path,lee2010efficient,peper2013brownian,lee2016brownian,utsumi2022computation}. In this context, we will specifically examine the thermodynamic properties of the BWC using a simplistic model proposed in~\cite{norton2013brownian}. The discussion begins with a concise overview of the expansion of a single-molecule gas. Subsequently, Brownian computers with different constraints are explored within this expansion model. The analysis leads to the inference that Bennett's assertion regarding the thermodynamic reversibility for the operation of BWCs is not tenable.

\begin{figure}[h]
  \includegraphics[width=0.85\columnwidth]{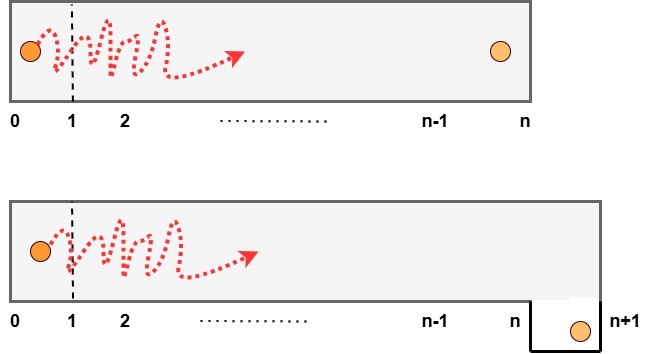}
  \caption{ A schematic of the BWC:  a) without energy trap, b) with energy trap.}
  \label{fig7}
  \end{figure}

\textbf{\textit{Single molecule gas expansion}:}
Let's contemplate an ideal single gas molecule situated at a specific temperature $T$ within a spacious chamber divided into $n$ parts by partitions, each with a volume $V$. In the initial phase, the gas molecule resides in the first cell with a volume $V$, as illustrated in Fig.~\ref{fig7}(a). Subsequently, the partitions are removed, allowing the single gas molecule to expand to a larger volume throughout the chamber.



The system Hamiltonian is $H = \mathbb{L} (p)$, where $p$ is the momentum of the molecule and $\mathbb{L}$ is a quadratic function of $p$. 
So, the entropy for the system is evaluated as $S = \frac{\partial}{\partial T} (k_BT\,\, \ln Z) = k_B \, \ln (nV) + C_p(T)$, where the contribution from the momentum perspective is included in the constant $C_p(T)$, \rr{$n$ denotes the total number of chambers}, and $Z$ is the partition function of the system. 


\textbf{\textit{Brownian computer}}:
From a thermodynamic perspective, a BWC can be likened to a single-molecule \rr{gas} expansion. In our discussion, we will differentiate between driven BWCs (where an external force propels the system) and undriven BWCs. Additionally, the introduction of a trap (a slight energy gradient to confine the molecule), as depicted in Fig.~\ref{fig7}(b), enhances the entropic force driving the system.

In the case of the \textit{undriven BWC}, it mirrors the single-molecule expansion, but its drawback lies in its lack of computational utility. The final equilibrium state in this scenario is uniformly distributed across all computation stages. Conversely, introducing a trap causes the molecule to be confined by the trap potential, resulting in a non-uniform final state. This alteration in the system increases its computational utility.

To accelerate the computational process, external energy (drive) needs to be supplied to the system. Among the considered configurations, the \textit{driven BWC} with the trap is currently the most resource-intensive in terms of thermodynamic \rr{(irreversible)} entropy required to propel the system.

In summary, from the thermodynamic analysis of the Brownian computer, it can be deduced that the $n$-chamber BWC is fundamentally a thermodynamically irreversible process, characterized by a minimum entropy amount of $k \, \, \ln n$. Introducing various parameters, such as an energy trap and external driving, to the system results in \rr{entropy production that renders the model irreversible}.


Bennett's misidentification of BWC as a reversible thermodynamic process can be attributed to the focus on tracking internal energy rather than thermodynamic entropy. Analyzing thermodynamic reversibility solely based on the tracking of internal energy is \rr{misleading}. The essential condition for verifying whether a process is reversible lies in tracking the total entropy of the system. If the total entropy ($S_{system} + S_{environment}$) remains constant throughout the process, it signifies a thermodynamically reversible process. Bennett and Landauer's oversight of not tracking the BWC's \rr{total} entropy resulted in the misidentification \rr{that the BWC model is a} thermodynamically reversible process. 


It has been indicated in recent studies~\cite{utsumi2022computation,utsumi2023thermodynamic} that the thermodynamic cost of executing a computational process becomes less significant when employing a token-based Brownian circuit for computational cycles. This stands in contrast to the case of logically reversible Brownian TM, where entropy production is directly proportional to the logarithmic function of the state space.

\section{Thermodynamics of computational models}\label{sec7}
The foundations of computer science are based on algorithms, data structures, and computation theory. In computer science, models of computation serve as mathematically precise frameworks for describing automated processes of symbolic reasoning. These models are diverse rather than singular, encompassing various approaches. A foundational concept in theoretical computer science is the existence of a class of models termed \textit{`Turing complete'} or `\textit{universal}'. These models exhibit two key properties: (i) mutual equivalence and (ii) broader generality compared to non-equivalent models. Equivalence implies that any computation representable within one model can be translated seamlessly into another, and vice versa. Turing-complete models possess the capability to \rr{execute} computations from non-Turing-complete models, but the reverse is not necessarily true.

Here we focus on the two primary aspects of computation: (a)
The finite automata/finite state machine (FA/FSM), which belongs to the class of non-universal models, and (b) the Turing machine
(TM) which belongs to the class of universal models. In the following, we briefly describe them. Subsequently, in the latter half, the thermodynamic aspects of these two computational models \rr{will be} discussed.


\subsection{Mathematical Foundations}\label{sec6}
The primary thermodynamic aspect of computation is the energetic cost of computation~\cite{maroney2009generalizing,faist2015minimal,parrondo2015thermodynamics,kolchinsky2017dependence,boyd2016identifying,ouldridge2017fundamental,boyd2018thermodynamics,wolpert2019stochastic,wolpert2020thermodynamics,riechers2021initial,riechers2021impossibility,kolchinsky2021dependence,kardecs2022inclusive}. Delineated by the enormous energetic cost of computation, the urge of considering a computational metric of success involving the resource cost of the computation has found \rr{renewed} attention of late~\cite{auffeves2022quantum}. 
The estimation of the thermodynamic cost of computation is based on the following axioms~\cite{li1992mathematical}: 

\vspace{.1in}
\noindent \textbf{Axiom 1:} No thermodynamic cost for a reversible computation process. \\
\textbf{Axiom 2:} Any irreversible process \rr{(irreversibly bit provided or deleted)} that occurs in a computation process has a thermodynamic cost.\\
\textbf{Axiom 3:} For a reversible computational process, where the input set $\eta$ is replaced by the output $\zeta$, the set $\eta$ ($\zeta$) is not provided (deleted) irreversibly. \\
\textbf{Axiom 4:} All physical computations are considered to be effective (i.e., it boils down to the formal notion of TM computation). 

Based on the axioms stated above the thermodynamic cost \cite{li1992mathematical} has been computed in terms of the computational complexity, called the \textit{Kolmogorov complexity} (KC)~\cite{li2008introduction,vitanyi2013conditional} of the bit string, \bb{ which quantifies the shortest possible description (or program) that can generate a given string using a TM. The thermodynamic cost of a computation is determined by counting the number of bits that are irreversibly provided or erased. This measurement accounts for information compression to ensure an optimal representation of the computational records.}

The  KC of computing a bit string $\zeta$ from the initial bit string $\eta$ is expressed as
\begin{equation}\label{Kolmog}
 K \bb{(\zeta \vert \eta )} = min\{\vert p_c\vert +\, \vert W_i\vert  : \xi_i(\bar p_c, \bar \zeta) = \eta, \, p_c \in \mathbb{S}^\star \}.   
\end{equation}

\bb{Here $p_c$ denotes the program which is a finite sequence of symbols (or a bit string) belonging to the set $\{0,1\}^*$, $W_i$ is the enumeration of the TM, and $|\bullet|$ is the cardinality of the bit string. The cardinality of the program $p_c$ is the length of the bit string representing the program $p_c$. The Turing machine indexed by $W_i$ computes $\zeta_i$. The enumeration of the partial recursive function defined as  $\xi_i(\bar p_c, \bar \zeta) = \eta$ is an effective invertible bijection from $N \times N$ to $N$, which effectively maps inputs (including the program and input bit string) to outputs
 (see Appendix \ref{AppendixB}). This formulation quantifies the minimal computational effort required to transform $\zeta$ into $\eta$, taking into account both program length and machine description length.}

\noindent \textbf{Theorem 1:}
The thermodynamic cost $E(\eta, \zeta) $ of computing $\zeta$ from $\eta$ is given by
\begin{equation}
E(\eta, \zeta) \approx  K (\zeta \vert \eta) + K(\eta \vert \zeta).
\end{equation} 
\rr{Outline of the proof of this theorem is provided in Appendix \ref{AppendixB}.}

\bb{The axioms introduced in~\cite{li1992mathematical} were designed to establish a framework for analyzing the thermodynamic cost of computational machines using KC. KC is a purely mathematical measure that does not consider the physical processes involved in executing a computation. Since it focuses solely on the minimal program length required to generate a specific output, it does not account for the energy cost associated with resetting a computational machine. As a result, the process of resetting has been deliberately excluded from the axiomatic formulation developed for this purpose.}

Furthermore, Zurek has shown~\cite{zurek1989thermodynamic} that the KC provides the energetic bound of individual computations. A generalized version of Zurek's bound has also been established~\cite{kolchinsky2023generalized}, which is applicable to all quantum as well as classical computations, including both stochastic and deterministic ones. The bound of the thermodynamic cost \bb{$E$} of computing $\zeta$ from $\eta$ reads:
\begin{equation}
 \frac{\beta}{\ln 2} \bb{E} \geq  K\bb{(\zeta \vert \eta )} -\log_2 \frac{1}{p(\zeta \vert \eta)} - K (\mathbf{P}) + \gamma_{ \mathcal{C}},
\end{equation}
where $\log_2 \frac{1}{p(\zeta \vert \eta)}$ is the noise associated with the computation, $K(\mathbf{P})$ denotes the KC of the protocol $P$, and $\gamma_{ \mathcal{C}}$ is the additive constant which depends on the universal computer $\mathcal{C}$ (independent of $P$). \bb{A practical physical setup, inspired by the widely used ``two-point measurement" schemes in quantum thermodynamics, is considered. In this framework, a computational subsystem $A$ executes the transformation $\eta \rightarrow \zeta$, while the subsystem $B$ serves as a bath. The protocol $P= (V, U,\beta,\epsilon)$ encompasses the chosen product basis $V$ for the two subsystems, the unitary $U$, the inverse temperature $\beta$, and the bath’s energy function $\epsilon$. The Kolmogorov complexity is calculated using the definition in Eq.~\eqref{Kolmog}.}

In~\cite{zurek1989thermodynamic}, Zurek conveyed that the loss of algorithmic information can be quantified in terms of KC of the shortest protocol possible. It can be considered as a \textit{``algorithmic fluctuation theorem"}  relating the second law of thermodynamics and the \textit{Physical Church-Turing thesis}~\cite{kolchinsky2023generalized}.


\subsection{Finite State Machine}

\subsubsection{Finite Automata: Basic Aspect}
\begin{figure}[h]
  \includegraphics[width=1.0\columnwidth]{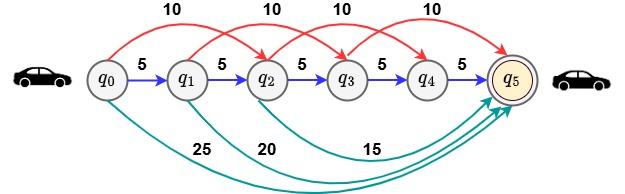}
  \caption{Transition diagram of the FA-controlled toll gate. 
$q_0$ is the start state. The state $q_5$ shown in two circles represents the final state (or the halt state). The insertion of the coins shown by arrows represents the alphabet.}
  \label{fig1}
  \end{figure}

First, consider a natural example of an automaton. Imagine a \textit{toll gate} controlled by a computer. Assume the gate remains closed until the required amount, say $25$ bucks, is paid. Moreover, assume that there are three sets of coins only: $5$, $10$, and $25$ bucks. Now, let us consider a situation where the driver of the vehicle inserts 25 bucks in the sequence (5, 5, 10, 5).  The state of the machines evolves as follows:
\begin{eqnarray}
q_0\rightarrow q_1\rightarrow q_2 \rightarrow q_4 \rightarrow q_5. \nonumber
\end{eqnarray}
The \textit{state diagram} with all possible combinations is shown in Fig.~\ref{fig1}. The gate opens or the computation ends if and only if the accepted (or halt) state (here $q_5$) is reached.
 Let us explain the steps of this controlled toll gate with a mathematical definition of FA below:

\noindent \textbf{Definition 1.} A FA is a 5-tuple $M = (\mathcal{Q},\, \Sigma,\, \delta,\, \bb{q_0}, \, F)$. Here,\\
1. $\mathcal{Q}$ represents a finite set whose elements are the states of the system.\\
2. $\Sigma$ also represents a finite set whose elements are called the alphabets (a finite set of symbols).  \\
3. $\delta: \,\, \mathcal{Q} \times \Sigma \rightarrow \mathcal{Q}$ represents the transition function. \\
4. $q_0 \in \mathcal{Q}$ represents the start state.\\
5. $F \subseteq \mathcal{Q}$ describes the set of accepting (or halt) states.

The transition function $\delta (q_i,b)$ takes an input $q_i \in Q$ and alphabet $b \in \Sigma$ (in Fig.~(\ref{fig1}) the sets of coins constitute the alphabet set $\Sigma$) and determines the output state $q_j \in Q$.
The machine has to record the state at any instant in time to determine the next step, i.e., to perform another transition or halt. A sequence of alphabets leading to halt states is called a word. The collection of the words forms a language of the FA, which happens to be a regular one (any finite language is called a \textit{regular language}) for FA.  

DFA means that the current state and the current symbol uniquely determine the next state.
NFA means that the same current state and the same current symbol may nondeterministically lead to different next states, but without any probability distribution. 
PFA means there is a conditional probability distribution over all possible next states given the current state and the current symbol.

FA have different forms like deterministic \bb{(DFA: the current state and the current symbol uniquely determine the next state)}, non-deterministic \bb{(NFA: the same current state and the same current symbol may nondeterministically lead to different next states, but without any probability distribution)}, and probabilistic finite automaton \bb{(PFA: there is a conditional probability distribution over all possible next states given the current state and the current symbol)}. In the stochastic \rr{(or probabilistic)} automaton, the single-valued transition function $\delta$ will be replaced by a conditional \rr{probability} distribution. One can also observe multiple accept states described in the literature as `terminal states' for the system~\cite{lawson2003finite}. One can even encounter multiple start states in the process. 

FA has a wide range of applications in computer science, like designing hardware, designing compilers, network protocols, and computation~\cite{lawson2003finite}. Furthermore, FA also has great impacts across different fields, including biology, mathematics, logic, linguistics, engineering, and even philosophy~\cite{bird1994one,baer1974automata,straubing2012finite}.  However, here we will focus on the computational aspects of FA and the role of thermodynamics in it.


\subsubsection{FA: Thermodynamic Aspect}


Though the reversible models of computation demonstrated by Bennett have vanishing thermodynamic costs following Axiom 1, they have major drawbacks for practical applications. All the reversible computation models take either infinite time to run or return a result with a very high probability of error. Recently, there has been exploration on the thermodynamic cost analysis of computational models in the \textit{quasi-static limit}~\cite{wolpert2015extending,strasberg2015thermodynamics,wolpert2023stochastic,gopalakrishnan2023push} that are inadequate to estimate the energetic cost of computations in real scenarios due to long execution time and huge error probability.  Therefore, the thermodynamic cost analysis of practical models of computation is of utmost interest. However, only very recently have there been a few explorations in this direction \cite{chu2018thermodynamically,ouldridge2022thermodynamics,manzano2024thermodynamics,ouldridge2023thermodynamics}.

Universal computation machine (aka TM, see details below) requires infinite tape, i.e., infinite memory at its disposal, which often makes the model unsuitable for practical applications. On the other hand, finite state machines (FSM), aka FA, albeit non-universal, that use finite resources, are an alternative model, as real-world computers have limited resources. The construction of a thermodynamically \rr{efficient} model~\cite{chu2018thermodynamically,ouldridge2022thermodynamics,kardecs2022inclusive} of FSMs has been in the limelight in recent times to estimate the energetic cost of such models. 

The appeal for studying FSMs offers an intriguing perspective: each state transition can be seen as a fundamental unit of computation, termed an elementary cycle. Understanding the thermodynamics of these elementary units enables one to grasp the thermodynamics of any FSM, as any FSM can be viewed as a sequence of these cycles. Moreover, these elementary cycles can be dissected further into more basic computational actions. Remarkably, only two types of basic computational steps are necessary for implementing any FSM: namely, the generalized versions of bit flips and bit sets. To grasp the thermodynamics of FSM,  in~\cite{chu2018thermodynamically} the authors constructed a thermodynamically consistent FSM model by designing FSM as a time-inhomogeneous Markov chain~\cite{wolpert2019space}. 

While investigating the energy consumption and the probability of accurate computation with the designed FSM model, Chu et. al. \cite{chu2018thermodynamically} were able to infer that in the high accuracy regime of the FSM, the probability of error scales polynomially, while the implementation cost, measured in terms of the work required for the cycle, increases only logarithmically. When expressed in terms of energy differences between states of FSM, it is observed that the average work scales linearly with $\Delta E$ (the energy difference), as expected, while the error decreases exponentially. Essentially, it indicates that the model can achieve perfect accuracy, albeit at the cost of infinite energy dissipation. However, quasi-deterministic computation with practically negligible error probabilities can be accomplished at a finite, even modest energy expense. Nevertheless, particularly for high accuracies, it is evident that the proposed model dissipates energy well beyond the theoretical limit.

Intriguingly, in the high accuracy limit, the size of the input alphabet and the size of the machine cease to significantly impact the cost of computation. One might speculate that a larger tape alphabet enables more information processing per computational step with only marginal increases in energy costs compared to smaller alphabets. This suggests that it may be more efficient to operate Markovian computers with larger alphabets rather than smaller ones. However, it's noteworthy that the error probability, in the worst-case scenario, can be observed to depend on both the size of the machine and the size of the alphabet. 

This observation \rr{conveys that for any algorithm, multiple FSM implementations exist, some favorable in accuracy and energy efficiency within an error tolerance, others less so, thus presenting} an intriguing possibility of trade-offs between performance, energy consumption, and accuracy, primarily determined by implementation rather than solely by the physical framework and the computation to be performed.

Following this direction, it has also been addressed~\cite{ouldridge2022thermodynamics}  that the cost for the computational characterization of the DFA divides \textit{regular language} into two classes. One is the \textit{invertible local map} and the other is the \textit{non-invertible local map}. In the former case, zero minimal cost is encountered, whereas in the latter case, a high cost is encountered.

An alternative approach other than the Markov chain model is addressed in~\cite{kardecs2022inclusive}, where they have developed a thermodynamic framework to define logical computers like DFA without specifying any extraneous parameters (like rate matrices, Hamiltonians, etc.) of the process that is considered to implement the computer. This framework doesn't require the entropy production to be zero and is derived from an exchange fluctuation theorem~\cite{crooks1999entropy,jarzynski2000hamiltonian,peliti2021stochastic,esposito2010three}. In particular, they use the \textit{Myhill-Nerode theorem}~\cite{lewis1998elements,hopcroft2001introduction} to prove that out of all DFAs which recognize the same language, the ``\textit{minimal complexity DFA}”  is the one with minimal entropy production for all dynamics and iterations.


\subsection{Turing Machine}\label{sec16}

\subsubsection{Turing Machine: Basic Aspect}

In 1936, Alan Turing proposed an abstract computation device~\cite{church1937turing}, later coined as the Turing Machine (TM), that can investigate the extents and limitations of all computable functions~\cite{hopcroft2000rotwani,savage1998models}. 
\textit{Church-Turing thesis}~\cite{copeland1997church} states that ``A function on the natural numbers is computable by a human being following an algorithm, ignoring resource limitations, if and only if it is computable by a \textit{Turing machine}." In the \textit{Physical Church-Turing thesis}~\cite{piccinini2011physical,cotogno2003hypercomputation}, it has been further modified on the physical ground that the set of functions which one can compute by utilizing the mechanical algorithmic \rr{methods} and abides by the laws of physics~\cite{pour1982noncomputability,moore1990unpredictability,wolpert2019stochastic,arrighi2019overview,wuthrich2015quantum}, are also computable with the help of TM\footnote{As all computational devices are physical, it has been argued in some works~\cite{baaz2011kurt,aaronson2005guest} that one might bring some restrictions to the foundation of physics by utilizing the properties of the TM.}. Various forms of definitions of the TM exist in the literature, which are computationally equivalent to each other. The formal definition of the TM is

\noindent \textbf{Definition 2.}  A TM is defined by 7-tuple ($\mathcal{Q}$, $\Lambda$, \bb{$\Sigma$}, $\delta$, $q_0$, \bb{$F$}, $q_r$). Here, \\
1. $\mathcal{Q}$ is a finite set that describes the non-empty set of states.\\
2. $\Lambda$ is a finite set depicting the input alphabets.\\
3. \bb{$\Sigma$} represents a finite set of tape alphabet and $\Lambda \,\subseteq \,\bb{\Sigma}$.\\
4. $\delta: \mathcal{Q}\times  \rr{\Sigma} \, \rightarrow \, \mathcal{Q} \times \rr{\Sigma} \times \{L,R, S\}$ is called the transition function. Here, $\{L, R, S\}$ describes the direction of the movement of the head of the tape. Based on the command, the head moves left, right, or stays in the same position on the tape.\\
5. $q_0$ ($q_0 \in \mathcal{Q}$) represents the start state of the Turing machine. \\
6. \bb{$F$} is called the accepted state or the halting state.\\
7.  $q_r$ is called the rejected state.

In other definitions of the TM, one can encounter multiple sets of halting states. 
 \begin{figure}[h]
  \includegraphics[width=1.0\columnwidth]{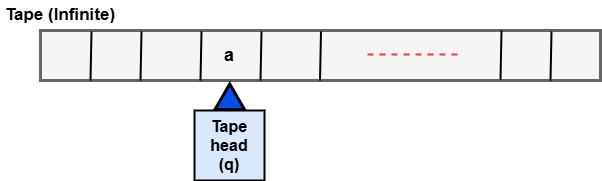}
  \caption{A schematic representation of a TM includes an infinite tape, where the current state is indicated by $q$. This tape is segmented into equally spaced square boxes, each containing symbols from the tape alphabet. The TM scans the tape using its tape head, which can move either left or right along the tape.}
  \label{fig2}
  \end{figure} 
At each step of the computation, the state of the TM reads the alphabet in the square where the tape head is placed and subsequently moves on to a new state $q'$ (see Fig.~\ref{fig2}). It writes a new alphabet ($a'$) on the tape, and then moves its tape head either to the left or to the right. This process is repeated until the system attains the accepted state. Mathematically, this map can be expressed as $(q, a) \rightarrow \, (q', a',  d)$, where $d$ denotes the right or left movement of the tape head.  For a given TM, the arguments of the transition states are called \textit{instantaneous descriptions} (IDs) of the TM. One can also encounter TMs with no halts. 

Earlier models like FA and push-down automata \footnote{The main difference of push-down automata with FA is that it has access to the top of the stack to decide which transition to take.} are not accurate in recognizing the language~\cite{arora2009computational}. On the contrary, the TM is considered to be the most accurate model.


TM has a great impact on the analysis of \textit{computational complexity}~\cite{hopcroft2000rotwani,moore2011nature,arora2009computational,sipser1996introduction,li2008introduction} and even in \textit{philosophy}~\cite{copeland2013computability}. One of the most important open problems in computer science that remains to be explored by the TM is whether $P = \, NP$~\cite{lipton2013people,razborov1994natural,fortnow2003p}. Limitations of mathematics, like \textit{G\"odel's incompleteness theorem}~\cite{godel1931monatshefte} still remain to be one of the main challenges to solve \rr{by TM.}


\subsubsection{TM: Thermodynamic Aspect}
TMs hold a central position in computation theory as a complete model of computation, unlike FSMs/FAs, which are non-universal models. This significance has prompted researchers to focus on the thermodynamic analysis of TMs in order to design computationally efficient models from a thermodynamic perspective.

For the thermodynamic analysis of TM, one needs to design a thermodynamically \rr{efficient} model of TM. Now, if one considers a single-tape TM, where the input tape is overwritten with the output tape, the computation becomes irreversible. Thus, to achieve reversibility in a TM, it's necessary to have at least two tapes—one for input and the other for output. In a reversible TM, it should be possible to retrace the computational path and retrieve the initial state of the TM. This requirement underscores the importance of maintaining the original information intact throughout the computation process.

A logically reversible TM was proposed by Bennett~\cite{bennett1982thermodynamics}, where he showed that a reversible TM needs at least four times the number of steps that are needed for the execution of a computation in an irreversible TM. Further generalization to Bennett's approach has been considered in~\cite{strasberg2015thermodynamics}. 
While Bennett considered a single input ($\mathcal{S}_{inp} \rightarrow \mathcal{S}_{out}$), the authors \rr{of}~\cite{strasberg2015thermodynamics} developed a TM which processes a continuous stream of input string on an infinite tape such as: ($\dots$, $a$, $\mathcal{S}_{inp}^{\prime}$, $a$, $\dots$, $a$, $\mathcal{S}_{inp}$, $a$, $\dots$). The input string denoted by $\mathcal{S}_{inp}^{\prime}$,  $\mathcal{S}_{inp}, \dots$ are separated by blank space symbols $a$, which denote the beginning and the end of the input string. Thus, the output string can be described as  ($\dots$, $a$, $\mathcal{S}_{out}^{\prime}= \mathbb{U}(\mathcal{S}_{inp}^{\prime})$, $a$, $\dots$, $a$, $\mathcal{S}_{out} = \mathbb{U}(\mathcal{S}_{inp})$, $a$, $\dots$), where $\mathbb{U}: \mathcal{S}_{inp} \rightarrow \mathcal{S}_{out}$.


The multiple-tapes TM model proposed in~\cite{strasberg2015thermodynamics} consists of four tapes: input, output, working, and history tape, respectively; and a computational cycle with five stages. The working and the history tape comprise the TM, while the input and the output are provided externally. The five stages of the computational cycle are: a) copy the input into the working tape, b) computation, c) copy the output into the output tape, d) retract the working tape to retrieve the input via the history tape, and finally e) erase the working tape.  


The dynamics of this logically reversible TM is modeled by a continuous Markov process that corresponds to a set of computational steps $\eta$ with probability $p_\eta$ that changes according to the $1^{\text{st}}$ order Markovian master equation~\cite{breuer2002theory,rivas2012open,rotter2015review}  
\begin{eqnarray} \nonumber \label{TRTM1}
\frac{d}{dt} p_{\eta} (t) & = & - \left( \mathcal{W}_{\eta+1,\eta} + \mathcal{W}_{\eta-1,\eta} \right) p_{\eta}(t)\\ 
& +  &  \mathcal{W}_{\eta,\eta+1} p_{\eta+1}(t) + \mathcal{W}_{\eta,\eta-1} p_{\eta-1}(t),
\end{eqnarray}
where $\eta \in \mathbb{Z}$, $\mathcal{W}$ is the rate matrix satisfying $\sum_{\eta^\prime} \mathcal{W}_{\eta,\eta^\prime} = 0 \quad \forall \eta$, and $\{\mathcal{W}_{\eta,\eta+1}, \mathcal{W}_{\eta,\eta-1}\}$ describe the forward and the reverse rate respectively, which obeys the detailed balance condition: $\frac{\mathcal{W}_{\eta,\eta^\prime}}{\mathcal{W}_{\eta^\prime,\eta}} = e^{-\beta (E_\eta -E_{\eta^\prime})}$. The rate matrix $\mathcal{W}$ is decomposed into blocks for each input during the computation. The transition between different blocks of the rate matrix is prohibited during the computation process.

For the thermodynamic cost analysis of this model, the physical system is associated with an energy landscape along the computational path~\cite{strasberg2015thermodynamics}, say, the logical and the successor states differ in energy by an amount of $\epsilon_c > 0$. 
The rate of entropy production for the computation is 
\bb{\begin{eqnarray}\label{TRTM3} \nonumber
\dot{S}(t) 
& =  & \frac{d}{dt} \mathcal{H}(t) + 2 \Gamma_c \, \beta \epsilon_c \, \sinh \left(\frac{\beta \epsilon_c}{2} \right) \geq 0,
\end{eqnarray}}
where $\Gamma_c$ denotes the rate setting for the overall time scale of the problem, \bb{$\mathcal{H}(t)= 1/2 \ln (4\pi \Gamma_c \epsilon_c \cosh{\beta \epsilon_c/2})$ is the Shannon entropy of the distribution.} 
In the limit, $\epsilon_c \rightarrow 0$, the rate of entropy production tends to zero, which confirms that this TM model works in a thermodynamically reversible manner in the steady-state regime. 

However, the small entropy production rate doesn't confirm that the overall entropy production is zero.  Like in the computational cycle, an unavoidable cost is encountered while resetting the TM. This is due to the increase in the Shannon entropy during the computation. This cost is dependent on the number of computational steps, and this verifies \textit{Norton's notion}~\cite{norton2013brownian,norton2014brownian} of thermodynamic irreversibility of computation.



Following various driving schemes~\cite{parrondo2015thermodynamics,van2013stochastic,esposito2010three}, a model to analyze the thermodynamic cost of the TM is proposed in~\cite{kolchinsky2020thermodynamic}. They have considered stochastic thermodynamics~\cite{wolpert2019stochastic} for the analysis of the dynamics of these physical processes. Interested readers can go through the review article~\cite{wolpert2019stochastic}, which provides a detailed analysis of the stochastic thermodynamics in different aspects of computation.
  
In~\cite{kolchinsky2020thermodynamic}, the authors propose a different approach to analyze the thermodynamics of TM by combining techniques from \textit{algorithmic information theory} and stochastic thermodynamics.  A discrete state system (equivalent to the input and output of the TM) is coupled to a reservoir at temperature $T$ and it evolves under the influence of the driving protocol. Three kinds of thermodynamic costs are encountered for this TM model:

(1) The heat generated during the execution of the realization of TM will be processed for each input $z$. It is denoted as $Q(z)$. 

(2) The heat generation for the entire computation that maps the input $z$ to the output $y$. This cost is referred to as the \textit{thermodynamic complexity} of $y$.

(3) \rr{The average heat $\langle Q\rangle$ produced by a TM realization, computed over the input distribution that minimizes entropy production.} 


Two physical processes are considered for the realization of the TM. 
The first physical process that is considered for the analysis is the \textit{coin-flipping} process for the universal Turing machine (UTM). This physical model is a thermodynamically reversible model, where the input is samples of the `\textit{coin-flipping}' distribution $p(z) \propto 2^{- l(z)}$, where $l(z)$ depicts the string length. The heat generated in this physical process is proportional to the computation program to execute the input $z$.


Being motivated by the \textit{physical Church-Turing thesis}, a semi-computable process (coined as \textit{domination realization}) is considered as the second physical process. It is shown that the second physical process is `optimal' in the sense that the heat generated by this process for any input $z$ is smaller than \rr{or equal to} any \rr{other} computable realization of TM on $z$.

The methods discussed thus far provide valuable insights into the thermodynamic aspect of physically realizing TMs. Considering the centrality of TMs to both physics and computer science, there exists a need for additional exploration to develop a more feasible and realistic model for the physical implementation of TMs.

\bb{\subsection{Quantum Computation: Energy cost}}

\bb{Thus far, we have explored the energy costs associated with traditional computational models. In today’s rapidly evolving digital economy, computing processes are consuming energy at an accelerating pace~\cite{liu2023potential,lannelongue2021green}. The widespread adoption of machine learning algorithms, large-scale language models, and \textit{data-intensive} operations has driven an unprecedented surge in energy demand~\cite{patterson2021carbon,arora2024sustainable,scholten2024assessing,an2023correspondence}. This escalating trend highlights the urgent need for energy-efficient alternatives to conventional computation. With the increasing deployment of large-scale AI models, energy consumption has become a critical concern in the modern computing industry. }

\bb{Given these challenges, the question arises: \textit{Could quantum computing provide a viable and energy-efficient alternative to classical computing?} Quantum computers, which leverage the principles of superposition and entanglement, have the potential to perform certain computational tasks exponentially faster than their classical counterparts. If harnessed effectively, quantum computing could significantly reduce the energy footprint of complex computations, offering a path toward a more sustainable and efficient computing paradigm, and is a subject of ongoing research~\cite{preskill2018quantum}.}

\bb{Quantum computation is widely expected to outperform classical computation across various computational resources. However, establishing a clear and definitive advantage in energy consumption remains an intricate challenge. This difficulty stems from the lack of a robust theoretical framework that directly correlates the physical concept of energy with the computational complexity of quantum algorithms. In classical computing, energy dissipation is inherently linked to irreversible operations, governed by LP. In contrast, quantum computing is fundamentally grounded in unitary evolution and reversible computation, making direct comparisons between the two paradigms highly nontrivial.}

\bb{Despite these challenges, recent advancements have made notable strides in bridging this gap. Researchers are actively pursuing both theoretical and experimental avenues to uncover the energy-efficiency benefits of quantum computing~\cite{meier2023energy,gois2024towards,green2022probing,pandit2022bounds,ikonen2017energy,martin2022energy,paler2022energy}. On the theoretical front, efforts are focused on formulating precise energy-complexity relationships for quantum algorithms, shedding light on the fundamental trade-offs between computational power and energy cost. It has been shown that quantum computing can offer a substantial amount of energy savings over classical methods for specific problems like-- Simon's problem~\cite{meier2023energy}, and the Fourier transform algorithm~\cite{gois2024towards}. Meanwhile, experimental investigations leverage state-of-the-art quantum processors, such as IBM’s quantum hardware~\cite{desdentado2021studying}, to empirically assess energy consumption in practical quantum computations. These studies seek to provide compelling evidence that, for specific computational tasks, quantum computers exhibit superior energy efficiency compared to their classical counterparts---bolstering the case for \textit{quantum supremacy} in the realm of energy-efficient computation.}


\vspace{4ex}

\section{Thermodynamics of Error Correction}\label{sec8}
During the communication \rr{or storage}, \rr{bits are} prone to noise, which tampers \rr{them}. \rr{Therefore}, the primary challenge in communication \rr{and storing}  is to detect these errors and reduce their influence on the \rr{information sent or stored}. Here, in this section, we explore the process to nullify the errors that occur during the communication \rr{or storage} process, both in classical as well as quantum regimes. 

In the quantum regime, the initial research was primarily focused on developing quantum codes~\cite{steane1996simple,steane1996multiple,knill1997theory,gottesman1998theory,gottesman1998fault,bennett1996mixed,knill1997theory,lieb1961two} that provided a rigorous framework for error correction~\cite{bennett1996mixed,knill1997theory,calderbank1998quantum}.  Now, \rr{ advanced} concepts like fault-tolerant quantum computation~\cite{shor1996fault,divincenzo1996fault,gottesman1998theory} provide the route-map to the threshold theorem for error correction in the quantum regime~\cite{knill1996threshold,aharonov1997fault}.

\subsection{Classical Error Correction}

In a communication process, the data is transmitted from the sender to the receiver end through a channel susceptible to noise, commonly referred to as a \textit{noisy channel}. The data string belongs to the set $\mathbb{S}$. The communication string undergoes encoding with the addition of extra bits (redundant bits). Upon reaching the receiver, the original message is reconstructed by processing the potentially corrupted message (due to a bit flip). This reconstruction process is known as \textit{decoding}.

\begin{figure}[h]
  \includegraphics[width=1.0\columnwidth]{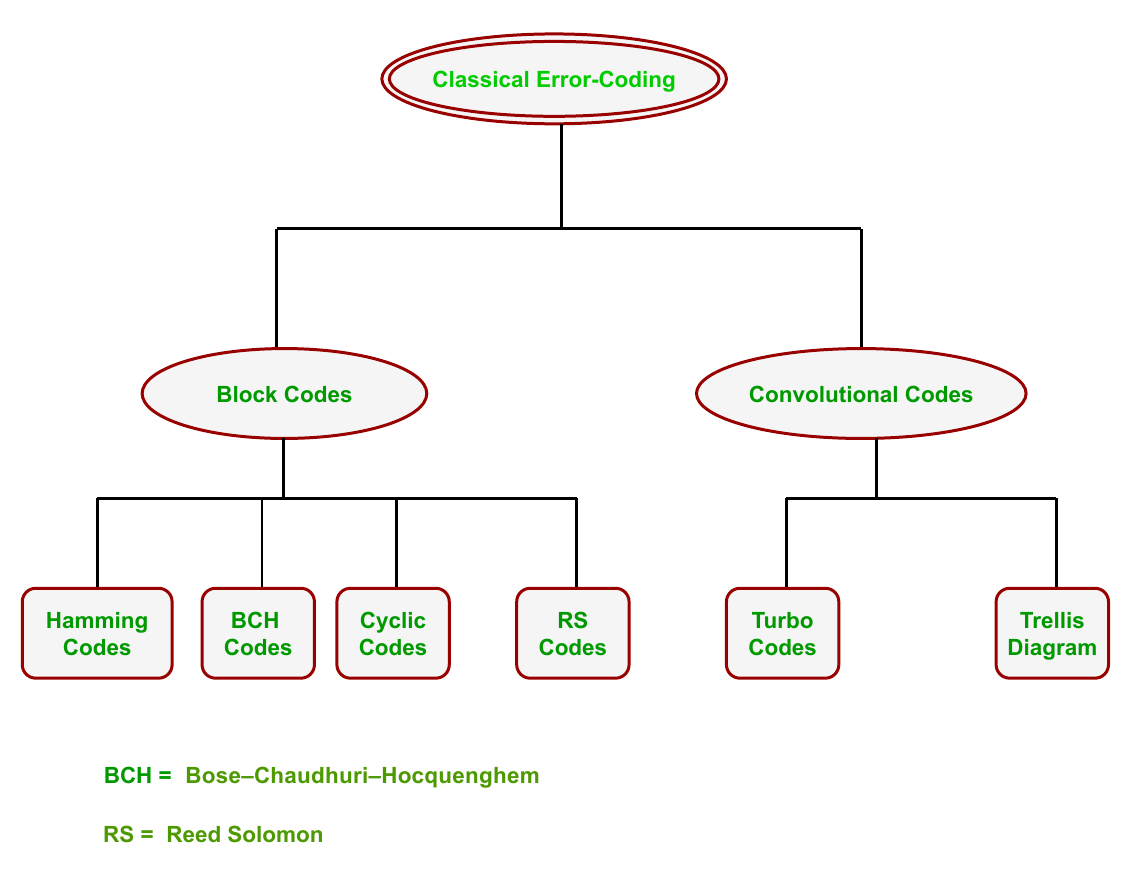}
  \caption{A schematic representation of the domain of the classical EC.}
  \label{figqu}
  \end{figure}

In the late 40's of the 20$^{\text{th}}$ century, the seminal work of Shannon~\cite{shannon1948mathematical}  led to the foundation of this field and was extended by Hamming in his work~\cite{hamming1950error}. Since then, this field has gained importance for developing better communication protocols.  The extent to which error correction (EC) of the missing bits is possible depends on the design of the error-correcting code (ECC). Generally, there exist two types of ECC, they are \textbf{block code}~\cite{jafarkhani2005space,adler1983algorithms,feltstrom2009braided} and \textbf{convolutional code}~\cite{dholakia1994introduction,forney1970convolutional,alfarano2023weighted} as depicted in Fig.~\ref{figqu}. 
Here we focus on a subfield of block code: Linear code. There are other models of error correction codes that are not covered here; interested readers can go through~\cite{pless1978fj,hoffman1991coding} for further information. 

The formal definition of the ECC is:\\
\textbf{Definition 3.} The ECC is defined as an \rr{injective} map from $n$ symbols (messages bits) to $m$ symbols (code bits):
\begin{equation}\nonumber
Enc: \Lambda_n \rightarrow \Lambda_m,
\end{equation}
where $\Lambda$ represents the set of symbols. \\

$\bullet$ The domain of the set, i.e., $\Lambda_n$ represents the message space, and $\Lambda_m$ represents the \rr{encoded message (codeword)}. Here $n$ denotes the message length.\\
$\bullet$\textit{ Block length}:  \rr{It refers to the length of the codeword} which is mapped to $m$-bit strings. \\
$\bullet$ \textit{Code}: \rr{A set of codewords produced by encoding messages.} In general, $m\geq n$.\\
$\bullet$ \textit{Rate}: It is defined as the ratio of $n$ over $m$. It \rr{quantifies} the efficiency of the \rr{code}.


A linear code is generally called a $[m,n,d]$ code, where $m$ describes the length of the codeword, $n$ denotes the length of the message string, and $d$ describes the minimum hamming distance. The hamming distance between two given vectors is \rr{given} by the number of positions the corresponding vectors differ.

\subsection{Quantum Error Correction}
Classical EC is a well-developed theory based on the demand for better communication systems. One-to-one mapping from classical to quantum error correction (QEC) is not possible as the quantum world has some constraints of its own, like qubits are \rr{governed} by the no-cloning principle~\cite{nielsen2002quantum}. \rr{As a consequence, even the simple repetition code\footnote{In the repetition code, the encoding is done as, for example, $0\rightarrow 000$ and $1\rightarrow 111$. It will correct a corrupted state, say $010$, to its majority value, $000$ in this case.} This belongs to the linear block code class and does not work in the quantum domain.}
\rr{Additionally, the phenomenon of wavefunction collapse upon measurement} makes quantum unique from classical. \rr{In} the seminal work of Peter Shor~\cite{shor1995scheme} the first QEC protocol was proposed. Shor in his work has demonstrated that quantum information can be encoded by exploiting the idea of entanglement of qubits. Works in this direction~\cite{calderbank1996good,preskill1998reliable,kitaev1997quantum,knill1998resilient,gottesman1998theory} have demonstrated that one can suppress the error rate in the quantum regime provided the qubits meet some physical conditions.  
\rr{A general state of a qubit is represented as} 
\begin{equation}\label{b1}
\vert \psi\rangle = \alpha \vert 0\rangle + \beta \vert 1\rangle,
\end{equation}
where $\alpha$ and $\beta$ represents complex number satisfying $\vert \alpha \vert^2 + \vert \beta \vert^2 = 1$. \rr{Thus, a qubit has the power to encode information in} the \rr{infinite number of possible} superposition of the computational basis states, which are denoted by  $\vert 0\rangle$ and  $\vert 1\rangle$. 
However, due to the digitization of the errors using the Pauli operator, the error counts are reduced to two fundamental errors~\cite{nielsen2002quantum}. One is the $X$-type error, which is the bit flip error (similar to classical), and the other is the $Z_{pau}$-type error, which is the phase error.
\begin{figure}[h]
  \includegraphics[width=1.0\columnwidth]{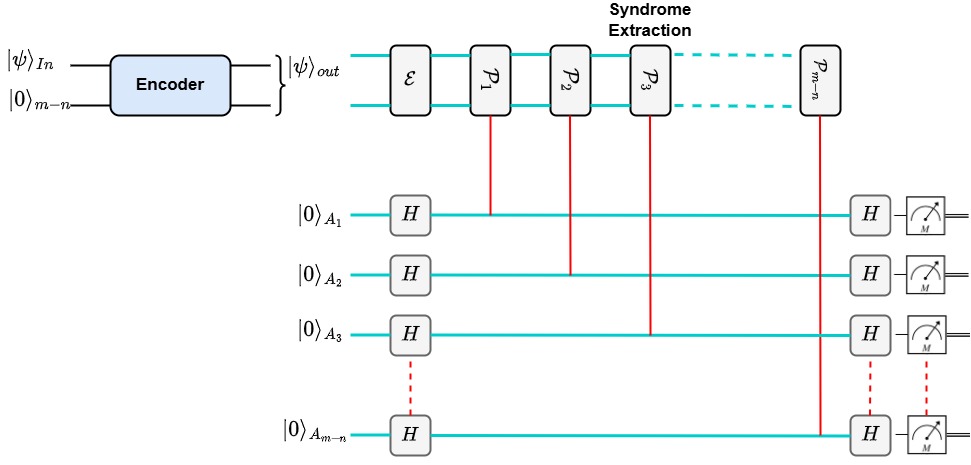}
  \caption{A schematic of the $[[m,n,d]]$  stabilizer code.  $\vert \psi\rangle_{In}$ represents the quantum data register, and $\vert 0\rangle_{m-n}$ represents the redundancy qubits. $H$ is the Hadamard operator. }
  \label{fig4}
  \end{figure}

Similar to the classical linear code, the \textit{stabilizer code} in quantum~\cite{steane1996error,calderbank1997quantum,calderbank1998quantum,gottesman1996class,gottesman1997stabilizer} is represented as $[[m,n,d]]$ (as shown in Fig.~\ref{fig4}).  Here $m$ represents the total count of \bb{physical} qubits, the count of the logical qubits is given by $n$, and $d$ describes the code distance \bb{which determines the number of correctable errors}. A stabilizer code encodes $n$ logical qubits into $m$ physical qubits. The stabilizer represents an abelian subgroup of the $m$-fold Pauli group. The notation of the quantum codes is in double brackets to differentiate it from the classical code, which is shown by a single bracket. 

\rr{Here, we described the basic intuition of QEC that is required in the latter half to understand the thermodynamic interpretation of EC.}
Interested readers \rr{who want to explore more about QEC} can go through the reviews~\cite{gottesman2010introduction,devitt2013quantum,lidar2013quantum,terhal2015quantum} in this direction, which covers QEC and its subfields.

\subsection{Thermodynamic Interpretation}\label{sec18}
From a thermodynamic perspective, EC is analogous to a refrigeration process. \rr{In the seminal work~\cite{vedral2000landauer}, Vedral has performed a thermodynamic analysis of EC both in classical and quantum domains, incorporating an MD-based model of EC. In~\cite{cafaro2014entropic}, the authors, building upon Vedral's result, have extended it for approximate QEC when the observation on the system is imperfect, implying sub-optimal information gain.} An alternative formalism is explored in~\cite{korepin2002thermodynamic} to establish the conditions of quantum codes and investigate QEC conditions from a thermodynamic standpoint.

\rr{In~\cite{vedral2000landauer}, the $``0"$ and $``1"$ states are represented by whether a single molecule of an ideal gas is in the LHS or RHS of a box, respectively, like Bennet's MD setup. Say, initially the molecule is in the LHS or $0$ state. If it expands isothermally, it works $\Delta W= k_B T \ln 2$ at the expense of the same amount of free energy stored in it. This increases its entropy by $\Delta S= k_B\, \ln 2$.}

\rr{If the molecule now jumps to the RHS, with probability $\frac{1}{2}$ say, then we say an error has occurred, which means the molecule has lost the power to perform work. To restore the work power we need to compress the molecule at one side of box and for the same we need to do $\Delta W= k_B T \ln 2$ amount of work.}

\begin{figure}[h]
  \includegraphics[width=0.7\columnwidth, height = 0.75\columnwidth]{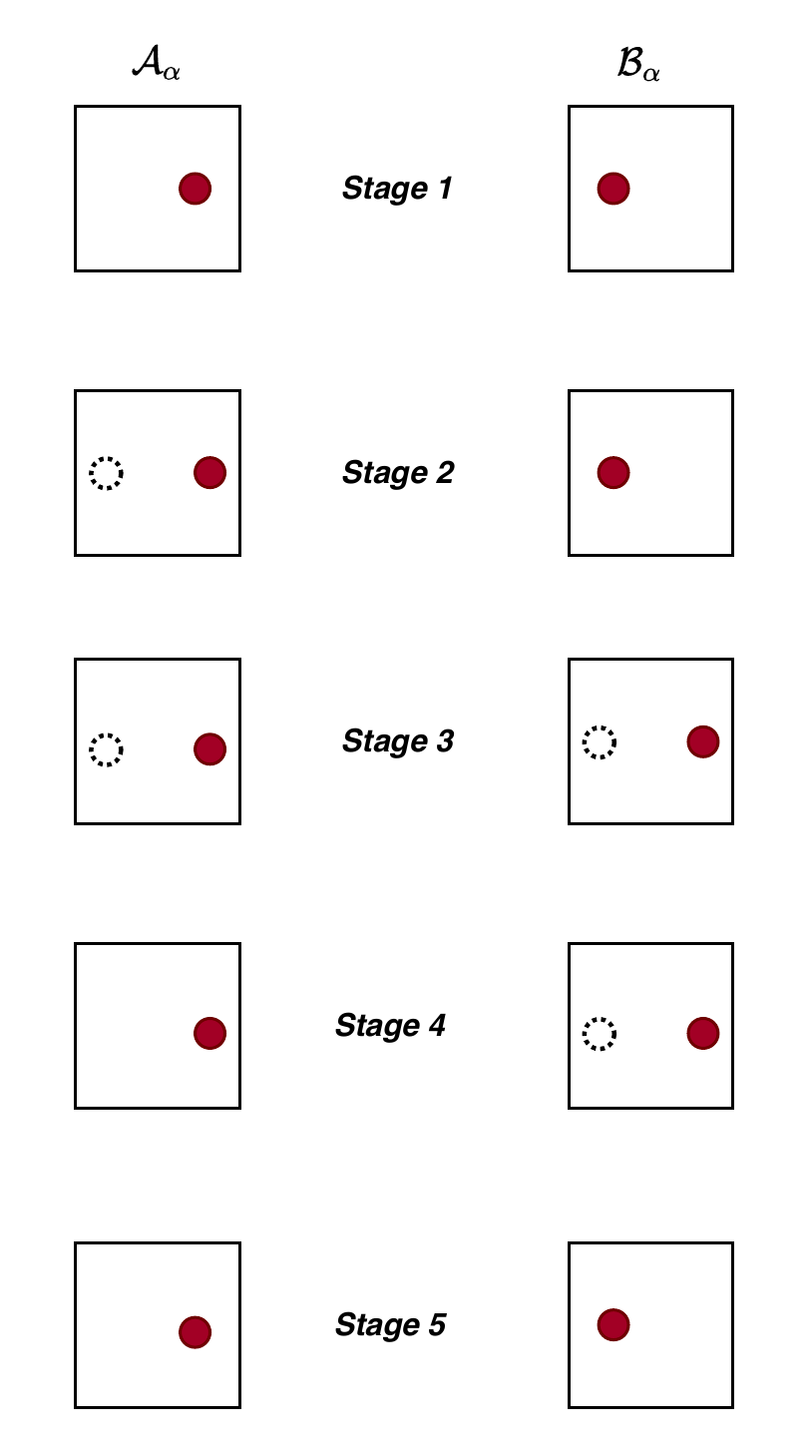}
  \caption{A schematic of the reversible cycle of the classical error correction protocol.} 
  \label{fig14}
  \end{figure}

\rr{Let us restore the work capacity of the molecule or correct the error with the help of another molecule} (as shown in Fig.~\ref{fig14})

(1) Let's consider that initially, the molecules are on the LHS and RHS of the respective boxes. 

(2) Now consider that some error occurs to the particle in the box $\mathcal{A}_\alpha$. 

(3) Now $\mathcal{B}_\alpha$ correlates itself (\rr{by some means whose details are not important}) with $\mathcal{A}_\alpha$ by observing it, such that the \rr{molecules} of both systems either occupy the LHS or the RHS of their respective boxes.  

(4) Based on the state of the system $\mathcal{B}_\alpha$ one will move the system $\mathcal{A}_\alpha$ to its respective side. This leaves the system $\mathcal{B}_\alpha$ in a randomized state.
\rr{Effectively we correct $A_\alpha$ by transferring the error to $B_\alpha$.}

(5) In the last step, the system $\mathcal{B}_\alpha$ is brought back to its initial state by isothermal compression, and \rr{it requires work.}

\rr{In summary, to correct $A_\alpha$ or reduce its entropy $B_\alpha$'s free energy is wasted just like how a refrigerator works. Now to restore $B_\alpha$'s free energy its entropy ($\Delta S= k_B\, \ln 2$) has to be dumped into the environment and it needs at least $\Delta W= k_B T \ln 2$ amount of work, which makes it exactly similar to the LP, and thereby consistent with the second law of thermodynamics.}


\rr{A similar} protocol of EC for pure states in the quantum domain is:

(1) \rr{The initial joint-state is given by  $ \vert \psi_{co}\rangle  \vert m\rangle  \vert env\rangle$, where $ \vert \psi_{co}\rangle$ describes the encoded state, $ \vert m\rangle$ represents the state of the measurement device, and $ \vert env\rangle$ is the  environmental state.}

(2) The combined-state after the introduction of the error \rr{by its environment} is described as $\sum_{j} Err_j  \vert \psi_{co}\rangle  \vert m\rangle  \vert env_{j}\rangle$, where $\{Err\}$ is the error operator and $ \vert env\rangle_j$ is the orthogonal environmental state. 

(3) Next, the environment is traced out \rr{leaving the combined state as} $\sum_j Err_j  \vert \psi_{co}\rangle \langle \psi_{co} \vert  Err_j^{\dagger} \otimes  \vert m\rangle \langle m  \vert $.

(4) \rr{Next, when} the system \rr{ is observed,} correlation is generated between the measurement apparatus and the system \rr{resulting the comnibed state} $\sum_j Err_j  \vert \psi_{co}\rangle \langle \psi_{co} \vert  Err_j^{\dagger} \otimes  \vert m_{j}\rangle \langle m_{j}  \vert $, where $\langle m_j  \vert  m_k \rangle = \delta(j, k)$. It \rr{means} the observation is perfect, \rr{otherwise, perfect recovery of the system would not have been possible}.


(5) 
\rr{In this step,} the correction of the error is executed. The state of the combined-system is $  \vert \psi_{co}\rangle \langle \psi_{co} \vert  \otimes \sum_j  \vert m_{j}\rangle \langle m_{j}  \vert$. 
Note that the state is not equivalent to the initial state, \rr{as} a resetting operation \rr{of the measuring device} is required. 

(6) To \rr{ reset it}, a garbage system is included \rr{giving the joint state} $  \vert  \psi_{co}\rangle \langle \psi_{co} \vert   \otimes \sum_j  \vert m_{j}\rangle \langle m_{j}  \vert  \otimes  \vert m\rangle \langle m  \vert$. \rr{Now,} by swapping the garbage and the measurement device, \rr{we get} $ \vert \psi_{co}\rangle \langle \psi_{co} \vert  \otimes  \vert m\rangle \langle m  \vert  \otimes \sum_j  \vert m_{j}\rangle \langle m_{j}  \vert $. Thus, the setup is reset and ready for another cycle of QEC.  

\rr{Note that the same entropy analysis as the classical EC is applicable here. The system, measuring device, and the garbage system play the role of system $A_\alpha$, $B_\alpha$, and the environment that sinks the entropy of $B$ respectively. This thermodynamic analysis of QEC can be extended even for a mixed state~\cite{vedral2000landauer}.}

If one considers \rr{imperfect measurement in step 4 of the above protocol} for the \rr{execution} of QEC, then \rr{only approximate recovery of the corrupted state is possible}. \rr{This problem is studied in detail in}~\cite{cafaro2014entropic}, \rr{where} the authors have introduced some ancilla qubits to keep track of the errors. \rr{They have argued that it is like an} refrigeration process, where the error \rr{(entropy) is transferred} from the data-qubits to the ancillary-qubits, \rr{thats}  \rr{purify} (cools down) the data-qubits. 



\rr{Recent studies have found} a similarity between the quantum heat engine and QEC codes~\cite{landi2020thermodynamic,danageozian2022thermodynamic}. The authors have fortified their intuition by conducting a comprehensive analysis of the thermodynamic properties of the quantum engine-based ECC.

\section{Miscellaneous} \label{sec9}
\rr{In this section, some complementary topics in the LP and thermodynamic aspects of computation are collated here.}

\subsection{Landauer Bound of Electronic Circuit}
In the context of electrical circuits, LP pertains to the process of resetting a bit in a digital memory element, such as a flip-flop or a register~\cite{lambson2011exploring,keyes1988miniaturization,hanggi1988bistability}. To overcome the thermal fluctuations that exist at finite temperatures, energy must be wasted to reset a bit to a reference state. 
This energy loss \rr{due to resetting} leads to an increase in the circuit’s temperature, potentially causing additional losses due to leakage current and other factors.  The LB of electronic circuits poses a significant constraint on the efficiency of digital computation and thus plays a crucial role in the design of low-power electrical circuits and the advancement of energy-efficient computing systems. 
Various methods~\cite{freitas2021stochastic,gopal2022large,freitas2022reliability} to minimize energy dissipation in electronic circuits have been explored lately. 

In~\cite{freitas2021stochastic}, a comprehensive theory is proposed for nonlinear electronic circuits affected by thermal noise. These circuits encompass devices with arbitrary I-V (current-voltage) curves but are subject to shot noise~\cite{sivre2019electronic,djukic2006shot}. This proposed theory incorporates a large class of electronic circuits, namely tunnel junctions, diodes, and metal-oxide semiconductor (MOS) transistors in sub-threshold operation. By considering the stochastic nonequilibrium thermodynamics of these circuits, the authors of~\cite{freitas2021stochastic}  formulate the thermodynamics of computing. The irreversible entropy production in such circuits is expressed in terms of thermodynamic potentials and forces. Specifically, the authors analyze a stochastic model of a subthreshold complementary metal-oxide-semiconductor (CMOS) inverter or NOT gate and derive an analytical solution for the steady state in the Markovian limit. In a nutshell, the investigation in~\cite{freitas2021stochastic} delves into how nonequilibrium thermal fluctuations impact the transfer function of the gate, utilizing the solution derived from the master equation.

\subsection{Landauer Bound in Switch Protocols}
Conventional CMOS technology faces the challenge of generating excessive heat during computation, far exceeding the theoretical \rr{bound}. This limitation poses a barrier to the advancement of switches. In pursuit of next-generation switches for advancing computer technology, the scientific community has turned its attention to \textit{mechanical switches}~\cite{jang2005nanoelectromechanical,cha2005fabrication,fujita20073,jang2008nems,jang2008mechanically}. 
In~\cite{neri2015reset}, the authors utilized molecular dynamics simulations to explore the minimum energy needed for reset and switch protocols in a bit encoded by compressed clamped-clamped graphene buckled ribbon.

Another alternative technology, bistable nanomagnetic switches, offers the ability to store information with low heat dissipation, where each logic state corresponds to a distinct equilibrium orientation of magnetization. 
In~\cite{madami2014micromagnetic},  the authors conducted virtual experiments based on quasistatic micromagnetic simulations at a fixed temperature in practical nanomagnetic switches~\cite{cowburn2000room,imre2006majority,csaba2002nanocomputing} to explore minimal energy consumption during a reset operation. Their findings confirm that the LB is accurately achieved for elliptical switches composed of elongated nanomagnets with lateral sizes below 100 nm, provided that the erasure technique employed is slow and occurs over an appropriate time interval. 

\subsection{Computer as a Heat Engine}
The study of the thermal machine is one of the primary perspectives of thermodynamics in classical~\cite{carnot1872reflexions,martini1983stirling,walker1985free,walker1985hybrid,van1985fundamentals,reed1898thermo,barton2019ericsson,scovil1959three,szilard1929entropieverminderung} and quantum regime~\cite{kim2011quantum,kosloff2013quantum,rossnagel2016single,martinez2016brownian,chattopadhyay2019relativistic,uzdin2015equivalence,chattopadhyay2021quantum,chattopadhyay2020non,mohan2024coherent,santos2023pt,mukhopadhyay2018quantum,sur2024many,das2019necessarily,naseem2020two,chattopadhyay2021bound,pandit2021non,singh2020optimal,singh2023asymmetric,PhysRevE.87.012140,GELBWASERKLIMOVSKY2015329,PhysRevE.106.054131}. In a simple sense, one can convey that the Carnot engine extracts an amount of heat $Q_{hot}$ from the hot reservoir at temperature $T_{hot}$ and transfers an amount of heat $Q_{cold}$ to the sink at temperature $T_{cold}$. The work done to execute this process is $W_{hot\rightarrow cold} = Q_{hot} - Q_{cold}$. Optimal efficiency is observed when
$\mathcal{L}_{hot}-\mathcal{L}_{cold} = 0$,
 where $\mathcal{L}_{hot} = - Q_{hot}/T_{hot}$ is the negentropy that is imprinted in the engine. So the loss that occurs in the execution of the process is just throwing away the negentropy of the amount $Q_{cold}/T_{cold}$. The concept of a computer being equivalent to a Carnot cycle has been explored~\cite{carnot1872reflexions,costa1989computer,brillouin1962science,prigogine1985self} and it is inferred from the results of~\cite{brillouin1962science,costa1989computer} that an ideal computer encounters a zero work balance, while the information delivered during a process is  $ \ln 2 \, dl= d\mathcal{L}_{hot} - d\mathcal{L}_{cold}$ (\rr{where $dl$ denotes the entropic cost of processing information}). \rr{ In contrast, while the ideal computer's operation suggests a loss of negentropy, it also emphasizes the potential for harnessing information to perform work, challenging traditional views of thermodynamic limitations. This duality underscores the complex relationship between information and energy in physical systems.}


A physical model has recently emerged, which focuses on autonomous quantum thermal machines for computational analysis~\cite{lipka2023thermodynamic}. These machines are comprised of interacting bits that are connected to baths with distinct temperatures, and  are referred as ``\textit{thermodynamic neurons}." In this setup, the machine undergoes evolution to a non-equilibrium steady state, and the computation output is determined by the temperature of an auxiliary finite-size reservoir. This model exhibits versatility and can be utilized to implement various linearly separable functions, like NOT and NOR gates.

\subsection{Nonergodic Systems and It's Thermodynamics}


For information erasure, Landauer argued that the system would confront a decrease in entropy while estimating the minimal dissipation by introducing an operation restore-to-one (RTO), whereas in~\cite{ishioka2001thermodynamics}, the authors have demonstrated that there is no change in the thermodynamic entropy even after the RTO operation. 

\begin{figure}[h]
  \includegraphics[width=0.7\columnwidth,height =0.6\columnwidth]{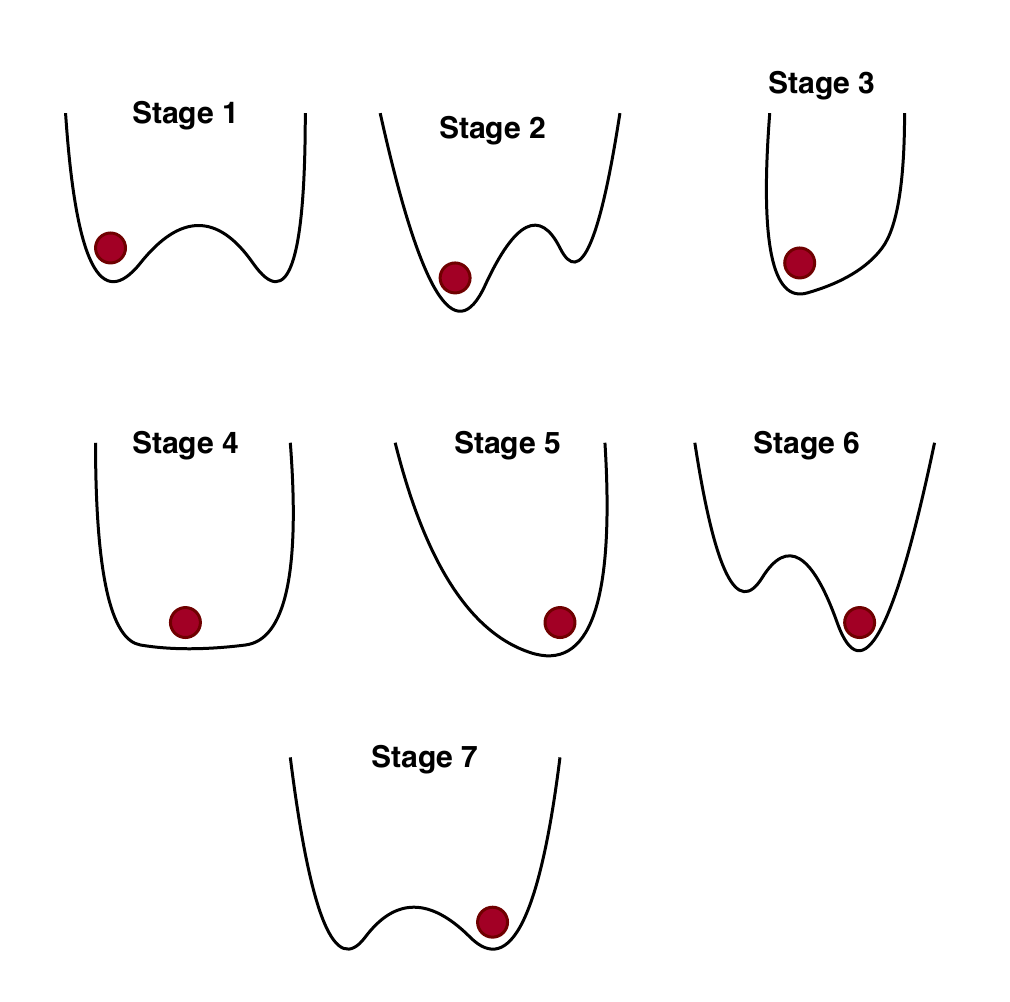}
  \caption{A schematic of the thought experiment is shown. The red solid sphere is the single atom of the system.}
  \label{fig15}
  \end{figure}

To support this assertion, the authors devised a thought experiment involving a particle confined in a bistable-monostable potential well interacting with a heat reservoir.  This model, termed \textit{quantum flux parametron}, was introduced by Goto et. al.~\cite{shimizu1989new,goto1996study}. In the analysis, the state of the system will be described as `1' when the particle is found on the RHS of the potential well and `0' when one finds the particle \rr{on the left.} The schematic representation of the thought experiment is depicted in Fig.~\ref{fig15}.








The first four steps of the thought experiment are called the \textit{erasing process}, and the final three steps represent the \textit{writing process}. If RTO is applied to the system, and the system is in the state \textit{zero}, the same configuration will be observed 
subjected to a condition that the state is known before the execution of this experiment. Landauer argued that the system will observe a decrease in entropy after RTO operation, but it is inferred  in~\cite{ishioka2001thermodynamics}  \rr{through the lens of Clausius’s definition of thermodynamic entropy that the erasure process characterized by a transition from a nonergodic to an ergodic state is fundamentally irreversible and entails the production of entropy. In contrast, heat generation predominantly occurs during the writing process. Remarkably, the reverse of erasure, namely, a transition from an ergodic to a nonergodic state, can be interpreted as a form of spontaneous symmetry breaking, which is associated with a decrease in thermodynamic entropy.} Thus, the thermodynamic entropy remains invariant during RTO operation. 

 \subsection{Thermodynamics of Algorithm}
It is widely believed that the emergence of quantum computers will aid in solving longstanding problems in number theory~\cite{borevich1986number,hua2012introduction}, combinatorial search~\cite{aigner1988combinatorial,katona1973combinatorial}, and even P and NP-classified problems~\cite{neukart2023thermodynamic}. 
To gain a deeper understanding of \textit{quantum speedups}, it is essential to examine a realistic model of computation that takes into account factors such as time complexity~\cite {sipser1996introduction} and time-space tradeoffs. Several works~\cite{banegas2018low,beals2013efficient,bernstein2009cost,fluhrer2017reassessing} have delved into these \rr{directions}. A recent study~\cite{perlner2017thermodynamic} in this direction has investigated quantum speedups from a thermodynamic perspective. For the analysis of algorithm\rr{ic} cost (in both classical and quantum regimes) from a thermodynamic viewpoint, the Brownian model of computation is employed. In~\cite{perlner2017thermodynamic}, the authors consider the collision-finding algorithm and preimage search for analysis in their thermodynamic interpretation of algorithms.

The parallel collision search algorithm proposed by Van Oorshot and Wiener~\cite{van1999parallel} stands as the leading classical collision finding algorithm. This algorithm can detect a collision in an expected serial depth of $\mathcal{O} (\sqrt{G}/G_{\alpha})$, where $G$ denotes the range of the function and $G_{\alpha}$ depicts the parallel processes with memory $\mathcal{O} (1)$.
Brassard, Hoyer, and Tapp (BHT) extended this algorithm in the quantum realm~\cite{brassard1998quantum}. The operations for this algorithm is $\mathcal{O} (G^{\frac{1}{3}})$ with memory size $G^{\frac{1}{3}}$. 

Giovanetti et. al. in their \rr{recent} study~\cite{giovannetti2008quantum}, address the memory cost by proposing a quantum random access memory (RAM) model, where the authors have conveyed that the memory access operation can be executed at logarithmic energy cost despite high gate complexity. So the question remains whether one can propose a realistic model where one encounters improvement in the complexity of the quantum algorithm over the classical one. The Brownian model is considered for the analysis, and it was inferred that the quantum algorithm has no advantage over the classical one. 

Applying a similar analysis \rr{to that used for the collision-finding algorithm can be envisioned for} \textit{Claw Finding problem}~\cite{tani2009claw,belovs2012span,jaques2019quantum,liu2023quantum,brassard1998quantum}, where the objective of the \textit{Claw Finding problem} is to find the collision between two functions with different domain sizes. The quantum version of this algorithm was investigated in~\cite{tani2007improved}, revealing that the energy cost for finding a collision \rr{is lower than the classical counterpart}.

The \textit{on-the-go erasure} protocol~\cite{meier2022thermodynamic} for the period finding algorithm has been addressed lately. 
Comparing Grover's algorithm~\cite{grover1996fast} with classical search algorithms rather than quantum versus classical collision search, reveals that Grover’s algorithm is more efficient \rr{in terms of query complexity} than its classical counterpart \rr{but both regime exhibit similar asymptotic energy consumption} (where powered and un-powered Brownian computation model is considered for the analysis). Grover's algorithm plays a substantial role in cryptography~\cite{grassl2016applying,lavor2003grover,hsu2003quantum,rahman2021grover,rahman2020quantum}, \rr{and thereby prompts us to explore the thermodynamic analysis of cryptographic protocols.}

\section{Conclusion and Future Direction}\label{sec10}

\rr{Both the thermodynamic and computational cost of}  a process have been of \rr{prime interests} for physicists as well as computer scientists~\cite{aifer2023thermodynamic,coles2023thermodynamic}. \rr{While the physicists are more concerned with the} thermodynamic cost \rr{a}, the computer scientists mostly care about  computational cost. The analysis of the thermodynamic computational cost for a process \rr{surfaced} from the \textit{physical Church-Turing thesis}~\cite{kolchinsky2023generalized}, where \rr{it was} conveyed that every computational process is physical. Various approaches have been considered for the analysis \rr{of thermodynamic cost} of the different computational processes. \rr{In this article we have reviewed the recent findings in these line in both the quantum and classical regime as well as provided an overview on the foundation and frameworks of thermodynamics of computation.} In recent times, the applications of these principles are not just restricted to the field of physics or computer science but are also prevalent in other fields like chemical networks~\cite{chen2014deterministic,soloveichik2008computation,murphy2018synthesizing,qian2011scaling}, molecular biology~\cite{prohaska2010innovation,benenson2012biomolecular}, and even in neurology~\cite{laughlin2001energy,balasubramanian2001metabolically}.

To get a better understanding of the bond between thermodynamics and the computational process, further investigations \rr{are still required.} For example, in the case of a finite automaton, \rr{it is tantalizing to} investigate the \rr{upper bound} thermodynamic cost that is required to accept a language for automata. Also, one can calculate the minimal cost for any deterministic finite automaton. Developing a theory to analyze the non-deterministic finite automata \rr{with thermodynamic tools could be promising}. Models to describe the complex Turing machine, and also network theory from the thermodynamic viewpoint, are open \rr{areas} of research.

Correction of errors during a computational process or communication is crucial. The analysis of the error correction protocols from a thermodynamic viewpoint is at its baby stage. Modeling of the EC models by the physical system to explain it thermodynamically needs further investigation.  So the thermodynamic approach to explaining EC is an open book to read.

\rr{While modern statistical and quantum information theory have advanced our understanding of the thermodynamic cost of computation, applying these insights to many-body physics, especially near quantum phase transitions (QPTs)~\cite{sachdev1999quantum,vojta2003quantum,osborne2002entanglement,heyl2018dynamical,sen2005dynamical,prabhu2011disorder} remains a significant challenge. QPTs in a quantum many-body system~\cite{bandyopadhyay2021driven,sur2020quantum,fetter2012quantum,tasaki2020physics,de2018genuine,PhysRevB.76.174303,PhysRevLett.91.207901,dutta2015quantum,gomez1996quantum,PhysRevLett.108.077206,fukuhara2013microscopic,PhysRevA.69.022311,PhysRevA.97.042330,PhysRevE.88.032906} occur at zero temperature due to quantum, rather than thermal fluctuations and are characterized by critical phenomena such as diverging correlation lengths and long-range entanglement. These features complicate the application of standard thermodynamic frameworks and raise open questions about how the minimal heat dissipation, governed by LP, behaves near criticality.  
Some compelling future direction involves relaxing the assumption of thermal reservoirs. These athermal characteristics could significantly modify the minimal energetic cost of erasure in general as well as near the QPT. Another fundamental challenge lies in understanding the behavior of LB as one approaches absolute zero, where the third law of thermodynamics and quantum effects become increasingly relevant. Additionally, pre-existing correlations between the system and a finite-sized reservoir may influence the erasure process in ways not captured by conventional formulations. Though there have been few explorations along these directions as we have reviewed, further explorations are essential for a deeper understanding of information thermodynamics in realistic, quantum-coherent environments.}

It is known that different systems have different heat signatures. One can utilize this property of the system for various purposes, such as for security in cryptographic protocols. So one can explore the communications protocols and \textit{crypto-systems} from a thermodynamic viewpoint. Algorithms in the form of a search algorithm from a thermodynamic viewpoint have already been analyzed. Further exploration in this direction is an open area of research. Thermodynamic analysis of quantum computations needs a rigorous investigation for a better understanding of quantum computers and to develop hardware with lower-cost functions.

\section{Acknowledgement}
P.C., A.M. would like to thank Nilakanta Meher and Saikat Sur of the Weizmann Institute of Science for their valuable inputs and suggestions. \rr{A.M. acknowledges partial support from the Anusandhan National Research Foundation, Government of India, under grant number ANRF/ECRG/2024/003836/PMS.}


%

\newpage
\section{Appendix}
\appendix

\section{\label{AppendixB}Theorem of thermodynamic computation}
\textbf{Theorem 1:} $E(\eta, \zeta) = K(\eta|\zeta) + K (\zeta|\eta)$ upto logarithmic term. 

\vspace{.2in} 
\textbf{Proof 1:} The upper and lower bounds for the thermodynamic cost are proposed. 

\textbf{\textit{Claim:}} $E(\eta, \zeta) \leq K (\eta|\zeta)+ K (\zeta|\eta) + 2[K( K (\zeta|\eta)|\zeta) + K (K (\eta|\zeta)|\eta)]$.

\vspace{.2in}
\textbf{\textit{Proof}:} The computation is divided in three parts. In the first part of the program it computes $\zeta$ from $\eta$, and is depicted by $p_c$ and $|p_c| = K (\zeta|\eta)$. In the second part of the program $q_c$ computes $ K(\eta|\zeta)$ from $\eta$ and $|q_c| = K (K (\eta|\zeta)|\eta)$, and in the final  part of the program $r_c$ computes $K (\zeta|\eta)$ from $\zeta$, and $|r_c| = K ( K (\zeta|\eta)|\zeta)$. Let's now analyze the computation process step by step. 

\begin{itemize}
\item In the first step  of the computation process $p_c$ computes $\zeta$ from $\eta$ and leaves behind garbage bits $g_a(\eta,\zeta)$.

\item $\zeta$ is copied, and then use one of its copies, along with the garbage bits, reverses the computation process to get $\eta$ and $p_c$. 

\item Now $\eta$ is copied, and then use one of its copies along with $q_c$ to compute $K (\eta|\zeta)$ along with the garbage bits. 

\item  The shortest program is executed, which is depicted as $s_c$ to compute $\eta$ from $\zeta$ with the help of $\eta$, $\zeta$, $K (\eta, \zeta)$. In this process, extra garbage bits are produced. 

\item Now, at this stage, $s_c$ is copied, and repeat the process shown in the third and fourth bullet. This helps to cancel the extra garbage bits. So we have $p_c$, $q_c$, $r_c$, $s_c$, $\eta$, $\zeta$. 

\item  $\zeta$ is copied again, and similarly, use one of its copies to compute $K (\zeta|\eta)$. It again results in garbage bits.  

\item The shortest program is executed but for $p_c$ to compute $\zeta$ from $\eta$ with the help  of $\eta$, $\zeta$, $K (\zeta, \eta)$. In this process, some extra garbage bits are produced.

\item Now, a copy of $p_c$ is deleted, and the process as shown in the sixth and seventh bullet is repeated. This helps to cancel the extra garbage bits. So we have $\zeta$, $r_c$, $s_c$, $q_c$. 

\item $\eta$ is computed from $s_c$ and $\zeta$ and then reduce a copy of $\eta$ by canceling it. Now we are left with $\zeta$, $r_c$, $s_c$, $q_c$. 

\item In the final step $s_c$, $r_c$, $q_c$ are erased. 
\end{itemize}

$s_c$, $q_c$, and $r_c$ are thermodynamically erased in this computational process, leaving behind the output $\zeta$. This provides proof of the claim. Now the second claim, i.e., the upper bound of the thermodynamic cost of the computational process, where $\zeta$ is computed from $\eta$. 

\textit{\textbf{Claim}:} $E (\eta, \zeta) \geq K (\zeta|\eta) + K (\eta|\zeta)$.

\vspace{.2in}

\textit{\textbf{Proof}:} The length of the shortest program to compute $\zeta$ from $\eta$ is defined as $K(\zeta|\eta)$. During the computation process,  garbage bits $g_c(\eta,\zeta)$ are produced. By definition~\cite{zurek1989thermodynamic}, the cardinality of the garbage bits is greater than or equal to the shortest program. So, to compute $\zeta$ from $\eta$, the garbage bits need to be erased,                   which is equivalent to at least $K (\eta|\zeta)$ bits. This proves the second claim. 

So the claims prove Theorem 1.

\end{document}